\documentclass[A4,twocolumn,amsmath,amssymb,rmp]{revtex4}

\usepackage{graphicx}
\usepackage{bm}
\usepackage{natbib}
\usepackage[T1]{fontenc}
\usepackage{times}
\newcommand{\araa}{Ann.\ Rev.\ Astron.\ \& Astrophys.}
\newcommand{\mnras}{Mon.\ Not.\ Roy.\ Astron.\ Soc.}
\newcommand{\aap}{Astron.\ \& Astrophys.}
\newcommand{\aaps}{Astron.\ \& Astrophys.\ Suppl.}
\newcommand{\apjl}{Astrophys.\ J.}
\newcommand{\apjs}{Astrophys.\ J.\ Supp.}
\newcommand{\apss}{Astrophys.\ Space Science}
\newcommand{\aj}{Astron.\ J.}
\newcommand{\pasj}{Proc.\ Astron.\ Soc.\ Japan}
\newcommand{\pasp}{Proc.\ Astron.\ Soc.\ Pac.}
\newcommand{\jqsrt}{J.\ Quant.\ Spectr.\ \& Rad.\ Trans.}

\newcommand{\msun}{M_{\odot}}

\newcommand{\rsun}{R_{\odot}}
\newcommand{\avir}{\alpha_{\rm vir}}

\newcommand{\sig}{\lower0.6ex\hbox{$\stackrel{\textstyle >}{\sim}$}\:}
\newcommand{\sil}{\lower0.6ex\hbox{$\stackrel{\textstyle <}{\sim}$}\:}
\newcommand{\sigs}{\lower0.4ex\hbox{$\stackrel{\scriptstyle
      >}{\scriptstyle \sim}$}\,}
\newcommand{\sils}{\lower0.4ex\hbox{$\stackrel{\scriptstyle
      <}{\scriptstyle \sim}$}\,}

\begin{document}

%RSK proposal for new title
%\title{Simulating Star Formation}
%MRK proposal for slightly modified version of Ralf's title
%\title{Modeling Stellar Birth -- On the Current State of Numerical Star-Formation Studies}
%\title{From Cores to Cores -- On the Current State of Numerical Star-Formation Studies}
%FH: ... and here's another version (just without ``On'' so it sounds less ``literary''.). 
%    Is the c2c thing somewhat too specific/specialised?
%\title{From Cores to Cores -- The Current State of Numerical Star-Formation Studies}
%RSK new attempt (as discussed via e-mail)...
\title{Numerical Star-Formation Studies -- A Status Report}

\author{Ralf S.\ Klessen}
\affiliation{Zentrum f{\"u}r Astronomie der Universit{\"a}t Heidelberg, Institut f{\"u}r Theoretische Astrophysik, Albert-Ueberle-Str.\ 2, 69120 Heidelberg, Germany}
\author{Mark R. Krumholz}
\affiliation{Astronomy Department, Interdisciplinary Sciences Building, University of California, Santa Cruz, CA 95064, U.S.A.}
\author{Fabian Heitsch}
\affiliation{Department of Astronomy, University of Michigan, 500 Church St, Ann Arbor, MI 48109, U.S.A.}

\date{\today}

\begin{abstract}
The formation of stars is a key process in astrophysics. Detailed knowledge of the physical mechanisms that govern stellar birth is a prerequisite for understanding the formation and evolution of our galactic home, the Milky Way. A theory of star formation is an essential part of any model for the origin of our solar system and of planets around other stars. Despite this pivotal importance, and despite many decades of research, our understanding of the processes that initiate and regulate star formation is still limited.

Stars are born in cold interstellar clouds of molecular hydrogen gas. Star formation in these clouds is governed by the complex interplay between the gravitational attraction in the gas and agents such as turbulence, magnetic fields, radiation and thermal pressure that resist compression. The competition between these processes determines both the locations at which young stars form and how much mass they ultimately accrete. It plays out over many orders of magnitude in space and time, ranging from galactic to stellar scales. In addition, star formation is a highly stochastic process in which rare and hard-to-predict events, such as the formation of very massive stars and the resulting feedback, can play a dominant role in determining the evolution of a star-forming cloud.

As a consequence of the wide range of scales and processes that control star formation, analytic models are usually restricted to highly idealized cases. These can yield insight, but the complexity of the problem means that they must be used in concert with large-scale numerical simulations. Here we summarize the state of modern star formation theory and review the recent advances in numerical simulation techniques.

\end{abstract}

\maketitle

\section{Introduction}
\label{sec:intro}

Stars are central to much of modern astronomy and astrophysics.  They are the visible building blocks of the cosmic structures around us, and thus are essential for our understanding of the universe and the physical processes that govern its evolution. At optical wavelengths almost all natural light we observe in the sky originates from stars. The Moon and the planets in our solar system reflect the light from our Sun, while virtually every other source of visible light further away is a star or collection of stars. Throughout the millenia, these objects have been the observational targets of traditional astronomy, and define the celestial landscape, the constellations.  
The most massive stars are very bright, they allow us to reach out to the far ends of the universe. For example, the most distant galaxies in the Hubble Ultra Deep Field are all characterized by vigorous high-mass star formation.  Understanding the origin of stars, at present and at early times, therefore is a prerequisite to understanding cosmic history.

Stars are also the primary source of chemical elements heavier than the hydrogen, helium, and lithium that were produced in the Big Bang. The Earth, for example, consists mostly of iron (32\%), oxygen (30\%), silicon (15\%), magnesium (14\%) and other heavy elements \cite{morgan80}. These are produced by nuclear fusion in the interior of stars, and enriching gas to the chemical composition observed today in our solar system must have required many cycles of stellar birth and death.

Today we also know that many stars harbor planetary systems around them, about 300 are known as of fall 2008 \cite{udry07}. The build-up of planets is intimately coupled to the formation of their host stars. Understanding the origin of our solar system and of planets around other stars has a profound impact on how we see our position in the universe. Questions whether we are alone, or whether there is life elsewhere in the cosmos are of broad interest to all of us.

Stars  and the planetary systems they may harbor are born in turbulent interstellar clouds of molecular hydrogen with a small fraction of dust mixed in. At optical wavelengths, we see these clouds as dark patches of obscuration along the band of the Milky Way. The dust component blocks the light from stars further away. At far-infrared,  sub-millimeter, and radio  wavelengths, however, the dust becomes increasingly transparent and we can look into these clouds. These observations reveal extremely complex morphological and kinematic structure, where patches of cold high-density gas are interspersed between regions of low-density warmer material \cite{ferriere01}. It is thought that this complicated texture is caused by supersonic turbulence that is generated by large-scale gravitational motions in the galaxy (such as spiral density waves) or by energy and momentum input from stars themselves \cite{maclow04,elmegreen04,scalo04}.

\begin{figure}[htbp]
\begin{center}
\includegraphics[width=\columnwidth]{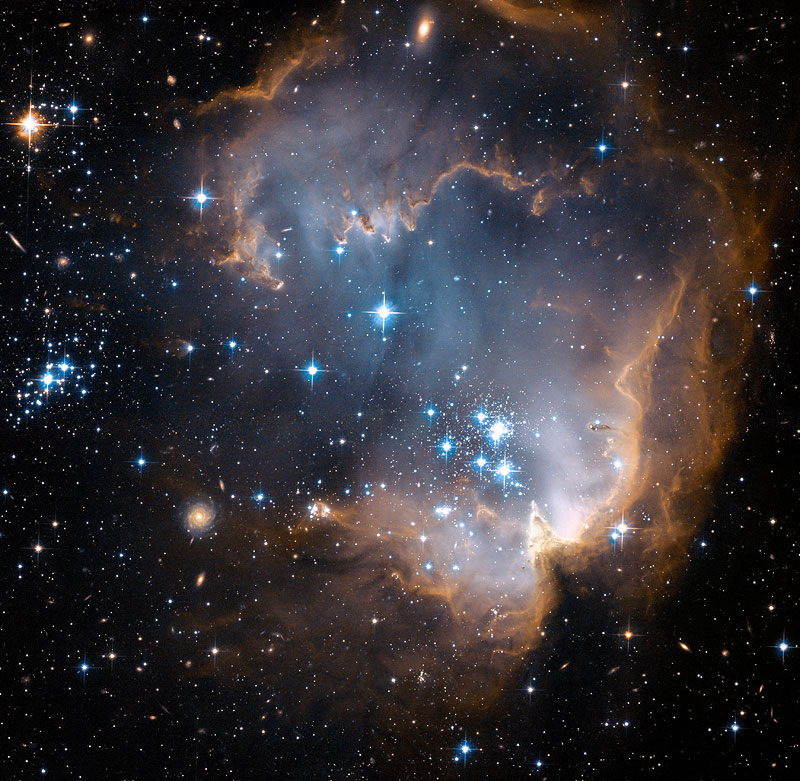}
\caption{Star forming region NGC~602 in the Large Magellanic Cloud, observed at optical and infrared wavelenths. The intense radiation from high-mass stars in the center of the young cluster has carved a cavity into the surrounding parental molecular cloud.  Elephant trunk-like dust pillars that point towards the hot blue stars are the 
%RSK3: removed tell-tale 
signs of this  eroding effect.  {\em (Image courtesy of NASA, ESA, and the Hubble Heritage Team)}}
\label{fig:NGC602}
\end{center}
\end{figure}

Within molecular clouds, supersonic turbulence and thermal instability lead to a transient, clumpy structure. Some of the resulting density fluctuations exceed the critical mass and density of gravitational stability. These clumps begin to collapse, their central density increases rapidly, and eventually they give birth to new protostars. Clusters of stars form in large regions that become unstable, within which contraction involves multiple collapsing cloud cores. A number of recent reviews have discussed various aspects of stellar birth in these clouds \cite{maclow04,Larson05,ballesteros07b,mckee07,zinnecker07}.

Stars and their parental clouds are connected via a number of feedback loops. Stars of all ages radiate and will thus heat up the gas in their vicinity. By doing so they influence subsequent star formation. Massive stars emit photons at ultraviolet wavelengths, creating bubbles of hot, ionized gas around them \cite{beuther07a,hoare07}, as illustrated in Figure \ref{fig:NGC602}. These so-called H\textsc{ii} regions are likely to quench further star formation in their interior, and thus set the star-formation efficiency in the region. The collective action of many H\textsc{ii} regions can destroy entire molecular clouds, and thus have the potential to influence the star-forming properties of galaxies on larger scales. The 
combined effect of large numbers of supernova explosions 
is another important mechanism for driving the supersonic turbulence ubiquitously observed in the galactic gas. By the same token, however, these feedback processes may trigger the birth of new stars. The very same processes that terminate star formation in one location may compress gas somewhere else in the galaxy, leading to new star formation. 

The density contrast between typical cloud densities and the hydrogen-burning centers of the final stars is enormous, about 24 orders of magnitude, and so is the corresponding spatial range (roughly 8 orders of magnitude). In addition to the large dynamical range, many different physical processes play a role at the various stages of the contraction process. 
On global scales we need to describe the formation of molecular clouds via large-scale flows of mostly atomic gas in a galactic disk. Internal turbulent compression in the star-forming cloud then sets the initial stage for the protostellar collapse of individual objects. The thermodynamics of the gas, and thus its ability to respond to external compression and consequently to go into collapse, depends on the balance between heating and cooling processes.  Magnetic fields and radiative processes also play an important role. Modeling star formation adequately  therefore requires the accurate and simultaneous treatment of many different physical processes over many different scales. 

In the past, progress has only been achievable by dividing the problem into smaller bits and pieces and by focusing on few physical processes or single scales only. Today, however, algorithmic advances and increasing computational power permit a more integrated approach to star formation. For the first time we are able to combine, for example, magnetohydrodynamics with chemical and radiative processes, and apply these numerical schemes to real astrophysical problems. It is this integrated view of stellar birth that is at the heart of this review. Our goal is to provide an overview of the recent advances in star formation theory with a special focus on the numerical aspects of the problem. We do not aim to be complete, for this we refer the reader to the recent reviews in this field \citep{maclow04,mckee07,zinnecker07}. Instead we will focus on three selected topics where we think numerical studies have had the largest impact and where we think our understanding of the physical processes that initiate and regulate stellar birth evolves most rapidly. We begin with the large scales and discuss the formation of molecular clouds in galactic disks and the numerical requirements and methodologies needed to do so consistently in Section \ref{s:molcloud}. We then zoom into individual star-forming regions and examine the transition from cloud cores to stars in Section \ref{sec:core-star}. As a third focus point, we discuss in Section \ref{sec:feedback} the effects of stellar feedback and examine how it alters the star formation process. Finally, in Section \ref{sec:outlook} we summarize and speculate about the future of numerical star formation research.

\section{The Sites of Star Formation: Molecular Clouds\label{s:molcloud}}

\subsection{Phenomenology of Molecular Clouds}
\label{subsec:phenomenology}

In regions of the interstellar medium (ISM) that are sufficiently dense and well-shielded against the dissociating effects of interstellar ultraviolet radiation, hydrogen atoms bind to form molecules. Star formation appears to occur exclusively within this molecular phase of the ISM. Molecular hydrogen is a homonuclear molecule, so its dipole moment vanishes and it radiates extremely weakly. Direct detection of cold interstellar H$_2$ is generally possibly only through ultraviolet absorption studies, such as those made by the the \textit{Copernicus} \citep{spitzer75} and \textit{Far Ultraviolet Spectroscopic Explorer (FUSE)} \citep{moos00, sahnow00} satellites. Due to atmospheric opacity these studies can only be done from space, and are limited to pencil-beam measurements of the absorption of light from bright stars or active galactic nuclei. Note that rotational and rovibrational emission lines from H$_2$ have also been detected in the infrared, both in the Milky Way and in other galaxies. However this emission comes from gas that has been strongly heated by shocks or radiation, and it traces only a small fraction of the total H$_2$ mass \citep[e.g.][]{vanderwerf00}. Due to these limitations, the most common tool for study of the molecular ISM is radio and sub-millimeter emission either from dust grains or from other molecules that tend to be found in the same locations as H$_2$. The most prominent of these is CO, although other tracers such as HCN are beginning to come into wide use.

\begin{figure}[tbp]
\begin{center}
\includegraphics[width=\columnwidth]{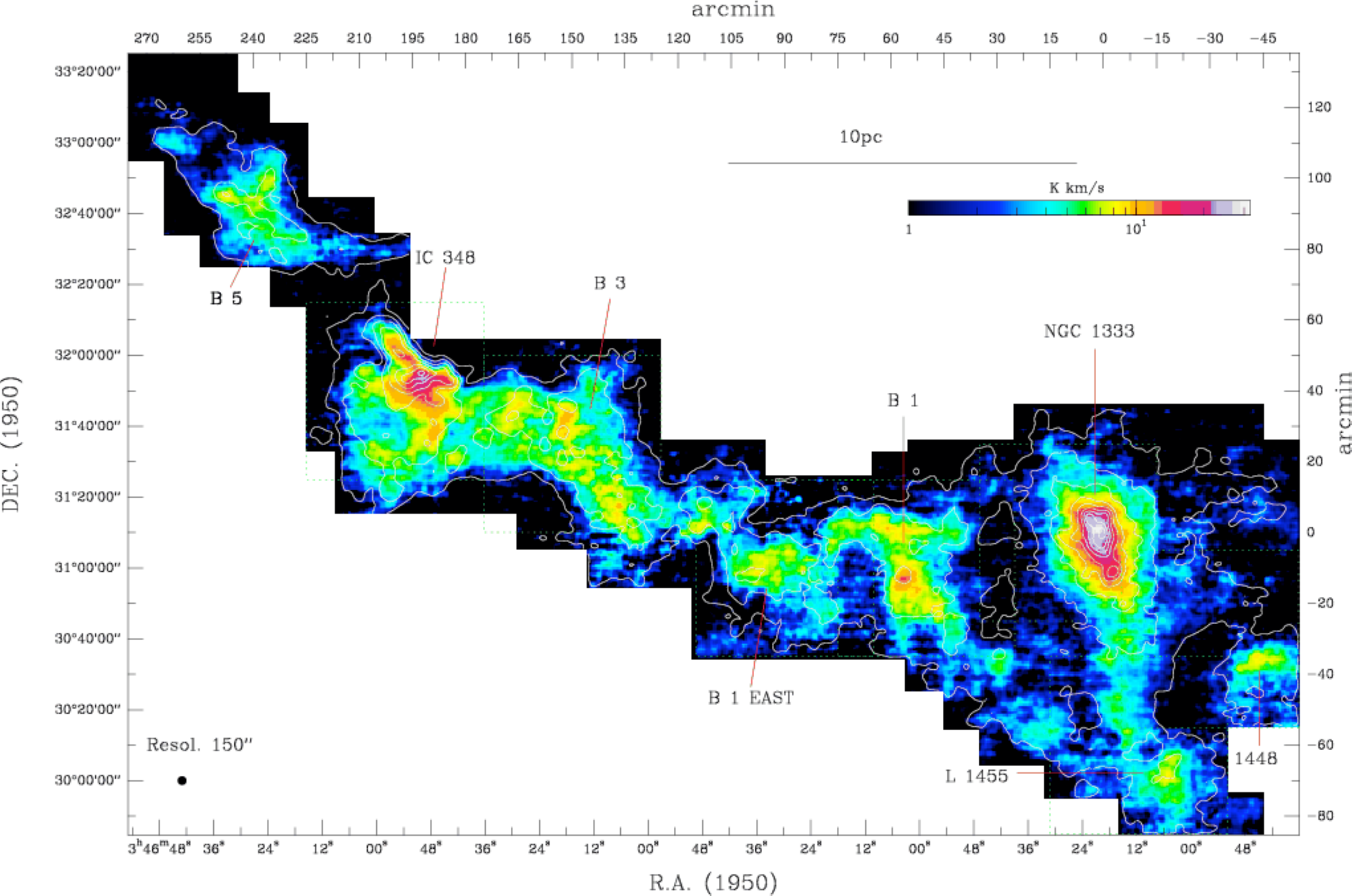}
\caption{Molecular cloud complex in the constellation Perseus. The image shows the distribution of CO line emission at radio wavelengths. This is a good tracer of total gas mass.
Clearly visible is the highly complex and filamentary morphological structure of the cloud.  {\em (Image from \citet{sun06}.)}}
\label{fig:perseus}
\end{center}
\end{figure}

As of this writing, the Milky Way and a few dozen Local Group galaxies have been mapped in the $J=1\rightarrow 0$ or $2\rightarrow 1$ rotational transitions of CO at a resolution better than 1$\,$kpc
\citep{dame87, solomon87, fukui99, engargiola03, rosolowsky03, rosolowsky05a, blitz07a, bolatto07a, fukui08a, walter08a, bigiel08, leroy08}, and a large number of more distant galaxies have been imaged at lower spatial resolution. The fractions of the ISM within the molecular phase in these galaxies ranges from no more than a few percent in low-surface density dwarfs to near unity in giant high-surface density systems. The highest molecular fractions are generally found in the parts of galactic disks with the highest total gas surface densities, but in the most actively star-forming galaxies the molecular fraction can reach $\sim 90\%$ even integrated over the entire galaxy \citep{mirabel89a, iono05a}. In all of the nearby galaxies where high resolution observations are possible the molecular gas is largely organized into giant clouds  (the so called giant molecular clouds or GMCs) of mass $\approx\,10^4-10^7\,\msun$ with average densities $\sim\,$100 H$_2$ molecules per cm$^{3}$, separated by a more diffuse intercloud medium. In the Milky Way and galaxies of lower density this medium is mostly atomic or ionized hydrogen, while in the densest nearby galaxies, such as M64 \citep{rosolowsky05a}, it is also molecular.

Molecular clouds across the Local Group all seem to display a number of properties in common. First, when studied with high spatial resolution clouds, they exhibit extremely complex and often filamentary structure, with column densities and corresponding 3-D densities that vary by many orders of magnitude (see Figure \ref{fig:perseus} or Table \ref{tab:MC-prop}). Nevertheless, when observed with low resolution, to within factors of a few all molecular clouds seem to have a similar mean surface density of $\sim 100$ $\msun$ pc$^{-2}$ corresponding to $0.035$ g cm$^{-2}$ \cite{heyer08, bolatto08a}. The constant surface density of molecular clouds is known as one of Larson's Laws \cite{larson81}, although there are a number of caveats with these relations and their interpretation \citep{ballesteros02}. Second, the clouds all display linewidths much greater than would be expected from thermal motion, given their inferred temperatures of $10-20$ K. The observed linewidth is related to the size of the cloud by
\begin{equation}
\sigma_{\rm 1D} = 0.5 \left(\frac{L}{1.0\mbox{ pc}}\right)^{0.5}\mbox{ km s}^{-1},
\end{equation}
where $\sigma_{\rm 1D}$ is the one-dimensional cloud velocity dispersion and $L$ is the cloud size \citep{solomon87, heyer04a, bolatto08a}. This is another one of Larson's Laws. 
These non-thermal linewidths have been interpreted as indicating the presence of supersonic turbulent motion, since both the low observed star formation rate (see below) and the absence of inverse P-Cygni line profiles indicates that they are not due to large-scale collapse. If one adopts this interpretation, then from these two observed relations one can directly deduce the third of Larson's Laws, which is that giant molecular clouds have virial parameters \citep{bertoldi92}
\begin{equation}
\avir \equiv \frac{5\sigma_{\rm 1D}^2 L}{G M} \approx 1,
\end{equation}
where $M$ is the cloud mass. This indicates that these clouds are marginally gravitationally bound, but with enough internal turbulence to at least temporarily prevent global collapse; whether they are truly in virial equilibrium is a topic that we discuss in detail below. (There is also a population of molecular clouds with virial ratios $\gg 1$, but these have masses $\ll 10^4$ $\msun$, and do not appear to host star formation \citep{heyer01}.) These relations appear to partially or fully break down in starburst galaxies with very high surface densities, where for example the molecular gas velocity dispersion can reach $\sim 100$ km s$^{-1}$ \citep{downes98}, but it is unknown whether there are analogous relations under these higher density conditions.

\begin{table}
{\caption{\label{tab:MC-prop}
 Physical properties of molecular cloud and cores$^a$}}
\begin{tabular}[t]{p{3.0cm}p{1.7cm}p{1.7cm}p{1.7cm}}
\tableline
& molecular cloud & cluster-forming clumps & protostellar cores \\
\tableline
Size (pc)                            &$2 - 20$   & $0.1-2$    &$\sil 0.1$\\
Density ($n({\rm H}_2)/{\rm cm}^3$)   &$10^2-10^4$& $10^3-10^5$&$>10^5$ \\
Mass (M$_{\odot}$)                  &$10^2-10^4$& $10 - 10^3$&$0.1-10$ \\
Temperature (K)                      &$10-30$    & $10-20$    &$7-12$ \\
Line width (km$\,$s$^{-1}$)           &$1 - 10$   & $0.3-3$    &$0.2-0.5$\\
Column density\\(g cm$^{-2}$) & $0.03$ & $0.03-1.0$ & $0.3-3$ \\
Crossing time (Myr)           & $2 - 10$   & $\sil 1$    &$0.1-0.5$\\
Free-fall time (Myr)           & $0.3 - 3$   & $0.1 - 1$    &$\sil 0.1$\\
Examples                             & Taurus, Ophiuchus & L1641, L1709 & B68, L1544\\
\tableline
\end{tabular}
{\footnotesize \vspace{0.2cm} $^a$~Adapted from \citet{Cernicharo1991} and \citet{bergin07}.}
\end{table}

The presence of supersonic turbulence in approximate virial balance with self-gravity indicates that in 
molecular clouds the turbulent and gravitational energy densities are of the same order of magnitude, and both greatly exceed the thermal energy density. If molecular clouds form in large-scale convergent flows (as we argue below in Section \ref{subsec:cloud-form}), then surface terms from ram pressure can also be significant and need to be considered  in the virial equations  \citep{Ballesteros06}.  There is also one further energy reservoir that we must mention, magnetic fields. The gas in molecular clouds is a weakly ionized plasma that is tied to magnetic field lines. Observations using Zeeman splitting \citep{crutcher99, troland08} and the Chandrasekhar-Fermi effect \citep{lai01, lai02} indicate that the field strength lies in the range from a few to a few tens of $\mu$G. The exact value varies from region to region, but in general the magnetic energy density appears comparable to the gravitational and turbulent energy densities. One can describe this state of affairs in terms of magnetic criticality. If the magnetic field threading a cloud is sufficiently strong, then it cannot undergo gravitational collapse no matter what external pressure is applied to it, as long as it is governed by ideal magnetohydrodynamics (MHD). A cloud in this state is called subcritical. In contrast, a weaker magnetic field can delay collapse, but can never prevent it, and a cloud with such a weak field is called supercritical \citep{1976ApJ...210..326M}. Observations indicate the molecular clouds are close to being, but not quite, magnetically subcritical \cite{2008arXiv0807.2862C}. For further discussions, see Section \ref{subsubsec:ind.cores}.

\subsection{Cloud Timescales and Cloud Formation}
\label{subsec:cloud-form}

\subsubsection{Characteristic Timescales for Molecular Clouds}
\label{subsubsec:timescales}

We can learn a great deal about molecular clouds by considering the timescales that govern their behavior. Because molecular clouds span a huge range of size and density scales, and their evolution times reflect this range, it is convenient to normalize all discussion of timescales to the free-fall time, defined as the time that a pressureless  sphere of gas with some initial starting mass density $\rho$ requires to collapse to infinite density under its own gravity: $t_{\rm ff} = \sqrt{3\pi/(32 G\rho)}$. For a cloud for which the virial parameter $\alpha_{\rm vir} \approx 1$, this is roughly half the cloud crossing time \cite{tan06a}, defined as the ratio of the characteristic size to the velocity dispersion $t_{\rm cr} = L/\sigma_{\rm 1D}$. The timescales $t_{\rm ff}$ or $t_{\rm cr}$ define the characteristic timescales on which behavior driven by gravity or limited by the internal gas signal speed can operate. For a molecular cloud detected via CO emission, with a mean number density $n\approx 100$ cm$^{-3}$, $t_{\rm ff} \approx 3\times 10^6$ yr. We now define some other useful timescales that can be determined from observations, and which yield strong constraints on how molecular clouds must behave. Any complete theory of star formation in molecular clouds must be able to explain each of the three timescales we describe.

\begin{figure}[htbp]
\begin{center}
\includegraphics[width=\columnwidth]{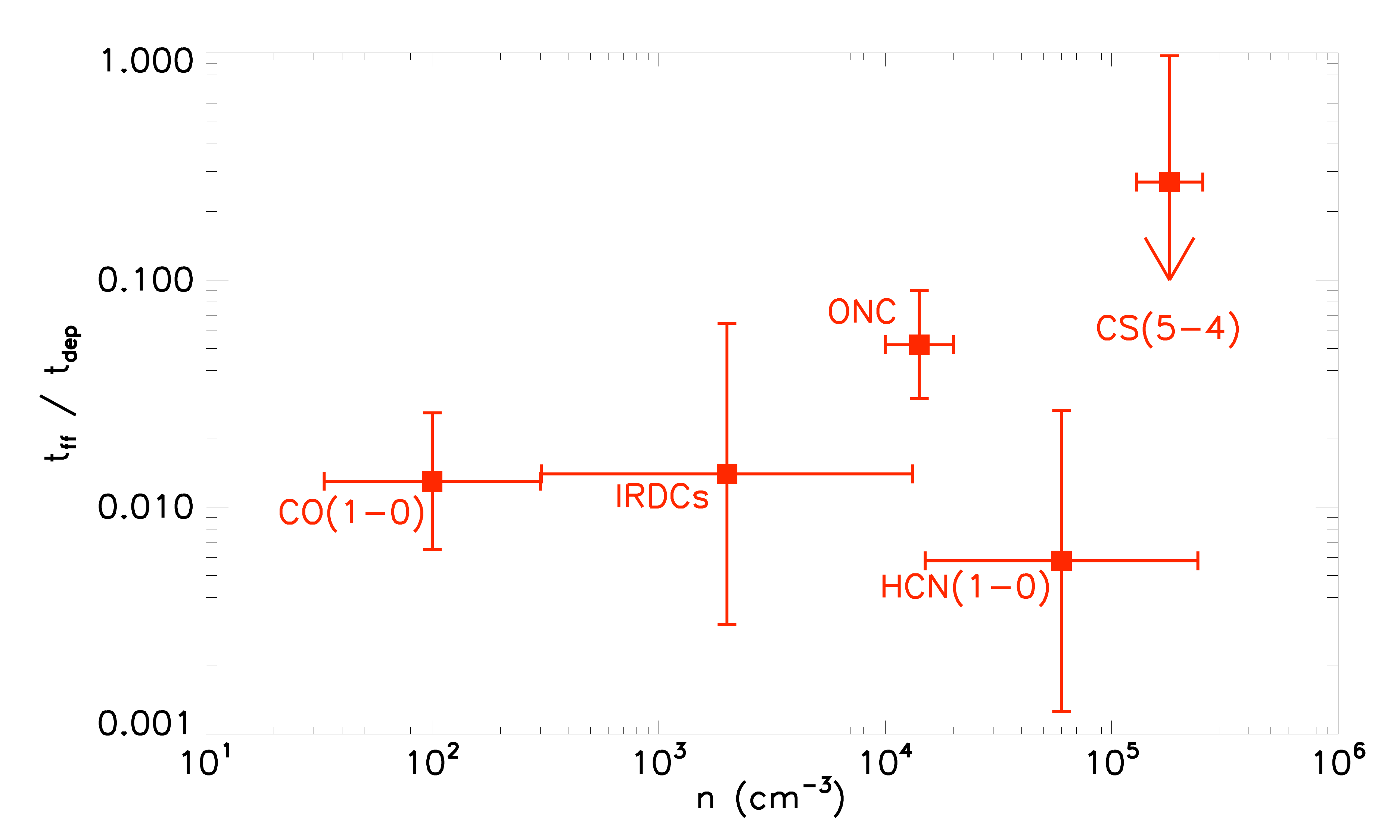}
\caption{Ratio of molecular cloud free-fall times to depletion times, versus the characteristic hydrogen density 
$n$ to which the indicated tracer of the molecular gas is sensitive. The depletion time is defined as the time that would be required to convert all of the gas into stars at the observed star formation rate. Each data point represents a survey of molecular clouds using a different tracer that probes gas of different densities, from CO (1$\rightarrow$0) emission to CS (5$\rightarrow$4) emission, which yields only an upper limit. IRDCs stands for infrared dark clouds, which are detected in infrared absorption, and ONC stands for the Orion Nebula Cluster, a single nearby star-forming region. 
{\em (Adapted from \citet{krumholz07e}.)}
}
\label{fig:sfeplot}
\end{center}
\end{figure}

\textit{The Gas Depletion Time.} Perhaps the most fundamental observational timescale for molecular clouds is the gas depletion time $t_{\rm dep}$, which is defined as the ratio of the mass of a molecular cloud (or population of clouds) to the star formation rate. This defines the time that would be required to convert the cloud completely into stars at its observed star formation rate (SFR), assuming this rate is constant over time. Estimating this number immediately yields a striking conclusion, which is perhaps the most basic problem in star formation. The disk of the Milky Way contains $\approx 10^9$~M$_\odot$ of molecular gas, and the observed star formation rate is only a few M$_\odot/$year, so the gas depletion time $t_{\rm dep}$ must be a few hundred Myr \cite{zuckerman74, mckee99}, roughly $100$ times the free-fall time. (Note, that if we compare this timescale with the age of the Milky Way of close to $10^{10}\,$yr, we conclude that a continuous inflow of fresh gas is required if the current SFR is at all representative and if we assume we are not living in times when the Milky Way is running out of gas soon. This problem gets worse if we consider proposals that the SFR was higher in the past \citep{MadauPozzettiDickinson1998}.) One can repeat this exercise for populations of molecular clouds in both the Milky Way and in other galaxies \cite{bigiel08}, using a variety of indicators of the star formation rate \citep{kennicutt98b}, and using a similarly wide variety of techniques to estimate masses of molecular clouds with various densities. Doing so yields the data shown in Figure \ref{fig:sfeplot}. In this Figure the $x$-axis indicates the characteristic density to which a particular method of measuring molecular gas is sensitive, and the $y$-axis shows the ratio $t_{\rm ff}/t_{\rm dep}$ for that gas. The fact that this ratio is $\approx1\,$\% for low density gas and either remains constant or slowly increases to at most $10\%$ at high densities indicates that the conversion of gas into stars must be inefficient or slow, in the sense that no more than a few percent of the total mass in molecular clouds in a galaxy can be converted into stars per free-fall time \citep{krumholz06d, krumholz07e}. This discrepancy is at the heart of any star formation theory, but before we can address it we must consider some other important timescales.

\textit{The Molecular Cloud Lifetime.} The gas depletion time tells us how long it would take to convert a molecular cloud into stars completely. The actual lifetime $t_{\rm life}$ of the cloud, however, is considerably shorter. Most of the cloud's mass is never converted into stars. Instead it participates in the perpetual cycle that connects the molecular, atomic, and ionized phases of the ISM \citep{ferriere01}. Molecular clouds form out of the atomic gas as discussed below (Section \ref{subsubsec:MCform}), convert some of their mass into stars, and then dissolve either by internal feedback (Section \ref{sec:feedback}) or large-scale dynamics. The total duration of this process is very hard to estimate.

In external galaxies, estimates of molecular cloud lifetimes are usually obtained from statistical relations between the location of molecular clouds and young star clusters. Star clusters can be age-dated, and determining at what ages they cease to be located preferentially close to molecular clouds gives an estimate of how long molecular clouds live once they form visible star clusters. Correcting for the population of molecular clouds that have not yet produced optically-visible clusters then gives an estimate of $t_{\rm life}$. In the LMC this is $t_{\rm life}\approx\,27$ Myr \cite{blitz07a} and in M33 it is $t_{\rm life}\approx\,20$ Myr \cite{engargiola03}, so $t_{\rm life}\sim 7-10\,t_{\rm ff}$, with a factor of $\sim\,$2 uncertainty. Due to the sensitivity limits of the observations, these estimates apply only to GMCs $\sim 10^5$ $\msun$ or larger. 

In star-forming regions within $\sim\,$500$\,$pc of the Sun, we can obtain ages estimates using individual young stars (as opposed to star clusters). Since stellar populations older than about $5\times10^6\,$yr are generally not associated with molecular gas anymore, this technique suggests molecular cloud lifetimes substantially shorter than $10^7$ years \citep{Hartmann01,ballesteros07a}. Placing stars on the Hertzsprung-Russell diagram results in stellar age spread from $1\,$Myr up to $3\,$Myr \cite{palla00, hartmann2003, huff06}, with considerable uncertainty but consistent with the above number. Unfortunately these nearby clouds have all masses well below $\sim\,$$10^5\,\msun$, so there is essentially no overlap between this population and the extragalactic one. Probably in part as a result of this selection effect, these regions are also denser than the giant clouds we can observe in external galaxies, and consequently have lower free-fall times, so $t_{\rm life} \sim 1-3\,$Myr corresponds to $t_{\rm life} \sim 1-10 t_{\rm ff}$ \cite{tan06a}.

\textit{The Lag Time.} The third timescale describing molecular clouds is the lag between when they form and when they begin to form stars, which we call the lag time $t_{\rm lag}$. For molecular clouds in the solar neighborhood (out to $800$~pc), 
the ratio of star-forming clouds over those without clear signs of star formation
ranges between $7$ and $14$  \citep{ballesteros07a}. Together with a median age of the
young stars in these regions of $1-2$ Myr \citep{pallastahler02,hartmann2003}
this entails a lag between cloud formation and onset of star formation of 
at most $t_{\rm lag} \approx 1$~Myr, i.e.\ stars begin to form immediately after (or even during)
the formation of the parental cloud. This is consistent with some extra-galactic observational evidence
that the spatial gap between spiral shocks in H~\textsc{i} gas and bright 24 $\mu$m emission downstream of the shock, presumably tracing star formation, corresponds to a lag time $t_{\rm lag} = 1 - 4\,$Myr \citep{tamburro07}. 

Before proceeding based on this conclusion, however, it is worth mentioning a caution. In both M33 \citep{engargiola03} and the LMC \citep{blitz07a}, the ratio of molecular clouds that have associated H~\textsc{ii} regions (detected either via H$\alpha$ or radio continuum emission) to those that do not is significantly smaller than the local ratio of star-forming clouds to non-star-forming ones: $3-4$ instead of $7-14$. This implies lag times of $\sim 7$ Myr, roughly $2-3t_{\rm ff}$, between GMC formation and H~\textsc{ii} region appearance. While the extragalactic clouds used for this measurement are much larger than the solar neighborhood ones and have free-fall times of $\sim 3$ Myr instead of $\sim 1$ Myr, they are comparable in size to the clouds probed by the geometric technique \citep{tamburro07}. The discrepancy between $t_{\rm lag}=2-3t_{\rm ff}$ and $t_{\rm lag} \sil 1 t_{\rm ff}$ between these techniques is therefore real. Its origin is unclear. One possibility that it is simply a result of the different criteria used to measure the onset of star formation: infrared emission versus H~\textsc{ii} regions. This possibility has yet to be quantitatively evaluated, however. In the absence of an explanation of the discrepancy, we tentatively conclude based on the IR data that $t_{\rm lag}$ is indeed short.
Because of this extremely short timelag, any property the molecular cloud has
to allow star formation, i.e.\ the strong density variations including small filling
factors, and the supersonic turbulence, must come from the formation process of the
cloud itself.

Strong density variations within molecular clouds are not only observed (Section \ref{subsubsec:statistics}, Table~\ref{tab:MC-prop}), they also are physically mandated. The free-fall time for a spherical cloud of uniform density does not depend on the radius. Thus if one neglects pressure forces, material at the edge of the cloud would arrive at the same time at the center as material close to the center, making it virtually impossible to form isolated stars. ``Distributed'' star formation can only occur if the cloud acquires high, localized density seeds during its formation process or similarly if pre-existing density fluctuations exist that become strongly amplified, such that the local contraction time is substantially smaller than the global one \citep{burkert04,Heitsch08a,Heitsch08c}.
The distribution of these high-density regions determines the degree of clustering of the resulting stars, with over-densities that are correlated on small length scales leading to isolated stars or small clusters, while over-densities correlated on large scales produce large clusters \citep{klessen00b,Klessen2001}. One possible explanation of the presence of small scale density inhomogeneity in newborn molecular clouds is that they form from atomic gas that is thermally bistable (Section \ref{subsubsec:MCform}). The onset of thermal instability leads to a wide-spread distribution of small-scale non-linear density fluctuations on very short timescales. This could explain how strong density variations occur on small scales even if molecular cloud turbulence is driven on large scales by the assembly of the cloud in the Galactic disk. In addition, GMC's are highly filamentary and show sheet-like morphology \citep{blitz07a}. This means that also boundary effects are important and any pre-existing initial fluctuations are more easily amplified compared to models mentioned above that assume spherical symmetry \citep{burkert04,hartmann07}. We note, however, that the smaller structures in their interior that form star clusters do appear to be more centrally concentrated \citep{plume97, mueller02, shirley03}.

\subsubsection{Molecular Cloud Formation}
\label{subsubsec:MCform}

The strict observational limits on $t_{\rm lag}$, the time between when molecules first appear and when star formation begins, have recently led
to a focus on the process of molecular cloud formation. This can be split into
three issues, namely the accumulation problem, the chemistry problem, and the issue of 
rapid fragmentation. We will discuss each of them in turn.

\textit{The Accumulation Problem}.
Building a molecular cloud requires assembling
a column density high enough for the dust to shield the cloud against the ambient UV-radiation,
and thus to allow CO formation. (H$_2$ forms earlier, because it self-shields very efficiently 
\citep{vandishoeck86, draine96, bergin04, krumholz08c}.
However, we ``see'' the cloud only once it contains CO.) Dust-shielding becomes efficient at column 
densities of approximately $N=2\times10^{21}\,$cm. If we were to accumulate our prospective
cloud at flow parameters typical for the inter-arm ISM, namely at densities $1\,$cm$^{-3}$ and
velocities of $10\,$km$\,$s$^{-1}$, it would take $\sim\,$60$\,$Myr to accumulate the shielding column density.
This timescale is too large, given the maximum lifetimes of GMCs of $\lesssim30\,$Myr 
(Section \ref{subsubsec:timescales}).

However, this seemingly compelling argument is not applicable, since it only addresses a one-dimensional
situation, resulting in a sky-filling molecular cloud. Various three-dimensional solutions have been 
suggested to allow molecular cloud formation within realistically short times. Probably the oldest 
relies on the Parker instability \citep{Parker1966}, a mechanism describing the buyoant rise of evactuated
parts of flux tubes in a stratified Galactic disk. The rising flux tube leads to material falling 
into the valleys, thus accumulating molecular clouds. In combination with a magnetohydrodynamical 
Rayleigh-Taylor instability to trigger the initial evacuation in spiral shocks \citep{Mouschoviasetal1974}, this
scenario predicts accumulation timescales of $30\,$Myr and typical cloud distances of $1\,$kpc along 
spiral arms. However, more recent numerical simulations of magnetized galactic disk gas dynamics \citep{kim06,DobbsPrice2008} 
only found weak signatures of a Parker instability acting along a spiral arm. Instead they
identified the Magneto-Jeans instability \citep{elmegreen87a,kim01} as a more dominant accumulation
mode in a magnetized disk. This instability is driven by in-plane magnetic fields countering the 
stabilizing effects of the Coriolis force, allowing self-gravitating contractions of overdense regions. 
Although we know that magnetic fields exist in galactic disks, it is also possible for molecular clouds to form on reasonable timescales in non-magnetized models provided the disk is globally gravitationally unstable \citep{li05b,li06,dobbs08b,tasker08,Tasker09a}.

On a more local scale, the interstellar medium is filled with flows of various types, many of which result
in piling up material through shocks. While it is not always easy to identify specific driving sources
in all cases (see, however, \citep{patel98,nigra08}), this is not surprising given
the complexity expected as a result of the interaction of neighboring flows.
In addition, the presence of massive molecular clouds well out of the Galactic plane 
(e.g., Orion) clearly suggests the need for some kind of driving.
Given the extensive impact that massive stars have on the interstellar medium --
H\textsc{ii} regions, stellar wind impacts, and ultimately supernova explosions -- it is difficult
to see how further creation of new star-forming locales by flows with scales of several to
tens of pc (or even kpc, in the case of spiral arms) could be avoided.
The notion of expanding shells piling up material has led to a ``collect \& collapse'' 
model \citep{elmegreen98}, connecting the formation of molecular clouds to energetic events in the ISM. 

\textit{The Chemical Timescale Problem}.
The formation of molecular hydrogen on dust grains -- the main branch under Galactic conditions -- is 
limited by three factors, namely the shielding from dissociating UV radiation, the dust temperature and the 
gas density \citep{tielens05}. In the extreme case of zero H$_2$ abundance in the assembling flows, 
dust shielding column densities need to be built up. However, due to its abundance of lines, 
H$_2$ strongly self-shields already at column densities of $N\approx\,10^{14}\,$cm$^{-2}$ 
\citep{vandishoeck86, draine96, tielens05}. 
Thus, alread small traces of H$_2$ in the accumulating flows will help to lower the timescales for molecule formation
\citep{Pringleetal2001}. 
The dust temperature determines the efficiency of H$_2$ formation on dust grains (i.e.\ what fraction of the accreted hydrogen atoms react to form H$_2$). The efficiency is of order unity for $T_{\rm dust} < 25\,$K, but drops sharply above that, although it may remain as high as $\sim 0.1$ even up to $T_{\rm dust} = 1000\,$K \citep{cazaux04}. 
The timescale to reach equilibrium between formation and dissociation is given by 
$\approx\,10^9/n$ yr \citep{HollenbachWernerSalpeter1971}.

Because of the complexity of the chemical reaction networks, most cloud chemistry models are restricted 
to one-dimensional geometries, albeit in different environments such as molecule formation behind shock
fronts \citep{bergin04} or in (close to) quiescent clouds \citep{goldsmith07} this is a good approximation. Models including hydrodynamical
effects such as shock compressions \citep{bergin04} or turbulence \citep{Glover07b} predict shorter
formation timescales -- on the order of a few $10^6$ years -- than static models \citep{Glover07a},
emphasizing the role of density variations and turbulent flows for the chemistry of the interstellar 
medium. 

\textit{The Fragmentation Problem}. 
The key to the observationally mandated rapid onset of star formation is to provide the parental cloud
with high-amplitude (non-linear) density perturbations during its formation. We will describe a 
scenario of flow-driven cloud formation and rapid star formation in the context of cloud formation in
spiral arms. While differing in details, other environments such as cloud formation in galaxy mergers or
in the Galactic molecular ring, are subject to similar physical constraints and may work along similar 
lines, although this is an issue still to be explored.

Rapid fragmentation of the accumulating flows results from the combined action of heating 
and cooling processes in the ISM. In the diffuse, warm interstellar gas, at densities of $n \approx1\,$cm$^{-3}$
photo-electric heating by dust grains dominates, while in denser, higher extinction gas, heating by cosmic 
rays is more significant, becoming dominant deep in the interior of molecular clouds. In the regime dominated by 
photo-electric  heating, the total heating rate per hydrogen nucleus varies by at most a factor of $10$ over
a range of densities from $n \approx 1$ to $10^3\,$cm$^{-3}$ \citep{wolfire95,wolfire03}. Cooling rates
below $10^4\,$K depend strongly on the abundances of heavy elements, but have only a weak dependence
on temperature at $T > 100\,$K \citep{dalgarno72}. Moreover, in the warm, diffuse ISM, the
energy radiated in the dominant cooling lines, such as the hyperfine-structure line of singly ionized carbon 
at $158\,\mu$m, scales as the $n^2$, while the heating rate depends (roughly) on $n$. Starting from an equilibrium 
situation, a small density increase thus leads to a cooling instability \citep{Field65}. If the
size of the density perturbation is small, with a sound-crossing time that is less than its cooling time, then
as the gas cools, its density will increase owing to compression from the surrounding warmer medium.
If the temperature dependence of the cooling rate is weak, then the increase in the cooling rate produced
by the growing density is greater than the decrease caused by the falling temperature. And so the
perturbation cools with ever faster growing density. This process will stop when the density dependence
of the cooling rate changes, for instance, if the level populations of the dominant coolants reach their
local thermodynamic equilibrium values. In this case the cooling rate scales only as $n$. It will also stop when the
temperature dependence of the cooling rate becomes steeper, as will naturally occur at low temperatures.
In the ISM, both effects are important, and the thermal instability vanishes once $n \sim 100\,$cm$^{-3}$ 
and $T \sim 50\,$K \citep{wolfire03}, resulting in a two-phase structure of interstellar gas, with a warm diffuse 
phase occupying large volumes, and a cold, dense phase with small filling factors in rough pressure equilibrium 
\citep{burkertlin00,heiles03}. 

This thermal instability (in its various forms) is at the heart of the rapid fragmentation of the accumulating flows. 
It has been studied in various contexts, such as
in generally turbulent media \citep{kritsuk02a,kritsuk02b,audit05,heitsch06b,robertson08},
in cloud formation behind shockfronts \citep{Koyama00,koyama02}, 
or in collisions of gas streams in spiral arm shocks or 
driven by e.g.\ expanding supernova shells \citep{audit05,vazquez07,Heitsch08a,hennebelle08b}. 
The strength of the thermal instability 
derives from a combination with dynamical instabilities breaking up coherent shock fronts
and shear-flow instabilities \citep{vishniac94,Vazquez06,heitsch06b},
and from the fact that its growth timescales are substantially shorter than those 
of the hydromagnetic and gravitational instabilities involved \citep{Heitsch08b}.  

\begin{figure}[htbp]
\begin{center}
\includegraphics[width=0.9\columnwidth]{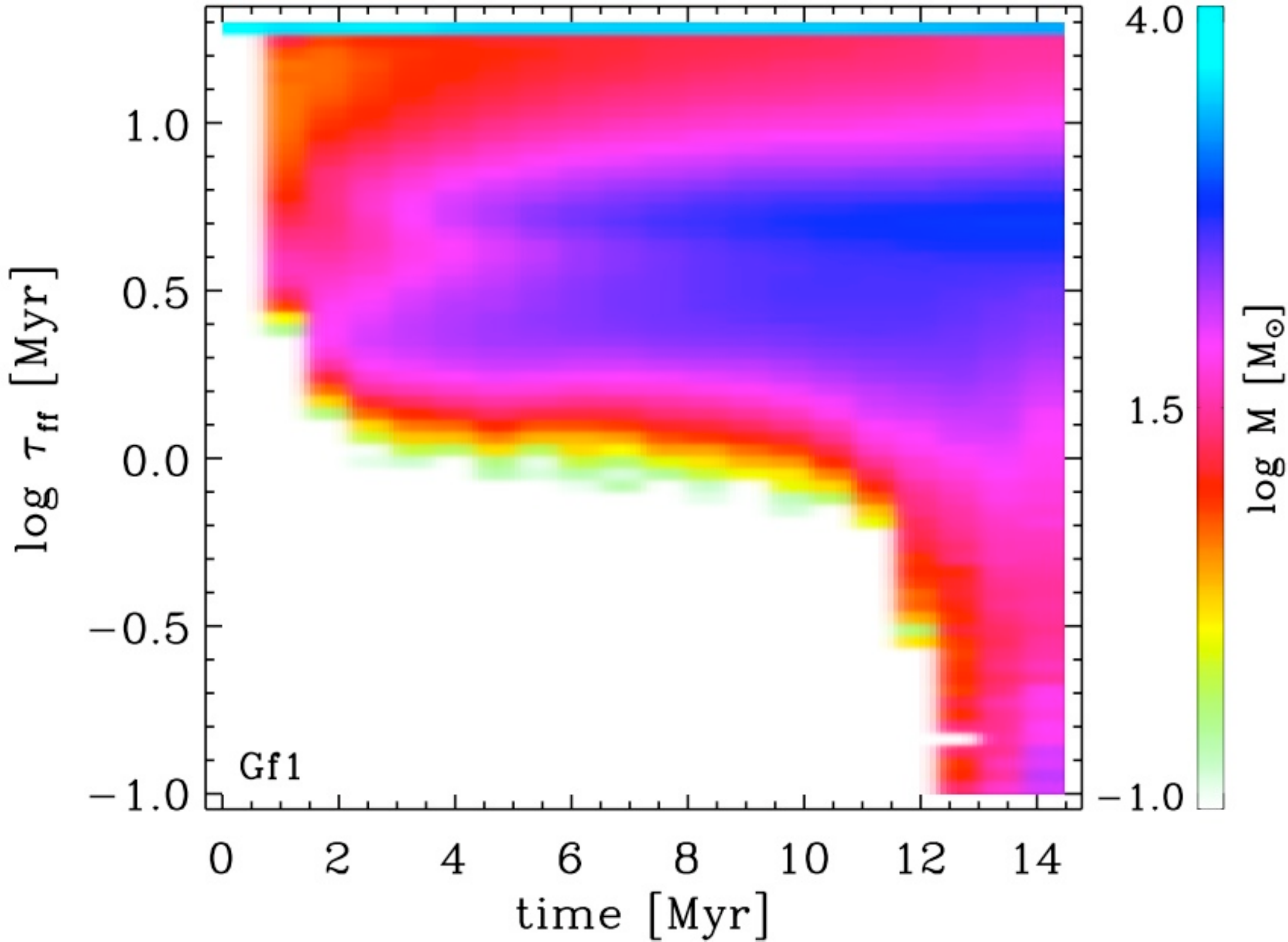}
\caption{Distribution of free-fall times against time in a molecular cloud forming in large-scale neutral
         hydrogen streams. Star formation sets in at $\approx\,11\,$Myr, when the local free-fall times
         get substantially shorter than those in the bulk of the cloud. For comparison, substantial
         CO is visible at $\approx\,10\,$Myr. 
         {\em (Adapted from \citet{Heitsch08c}.)}}
\label{f:freefalldist}
\end{center}
\end{figure}

The principal effect of this rapid thermal fragmentation is best gleaned from the evolution of the 
free-fall times in the forming cloud \citep{Heitsch08c}. Figure~\ref{f:freefalldist} shows the distribution of free-fall 
times against evolution time in a molecular cloud being formed by two colliding flows of warm neutral
hydrogen gas. Initially, the bulk of the cloud mass is at free-fall times longer than the simulation
duration. At around $7\,$Myr, a substantial mass fraction of the cloud has dropped to free-fall times 
as short as $3\,$Myr. Substantial CO has formed by $\approx$$\,10\,$Myr, while star formation sets in 
at $\approx$$\,11\,$Myr, when noticeable mass fractions appear at free-fall times substantially shorter 
than those in the bulk of the cloud.

One of the key realizations in contrast to earlier models
of turbulent fragmentation using periodic boxes is that the finite cloud geometry 
is crucial not only for the rapid onset of star formation \citep{vazquez07,Heitsch08a}, but also for the rapid formation
of CO to overcome the otherwise stringent limitations set by the inflow density and speed \citep{Heitsch08c}.
Thus these models bridge a gap between the large-scale simulations of Galactic disk dynamics discussed earlier, and
the detailed models of turbulent fragmentation to be discussed in Sec~\ref{sec:core-star}. The large-scale models do 
not have sufficient resolution to address the fragmentation and internal dynamics of the resulting molecular 
clouds, while models on smaller scales generally have to make simplifying assumptions about the boundary conditions. 

\textit{What's Missing in Cloud Formation Models?}
Perhaps the most serious complication with flow-driven cloud formation models is that by themselves they address only one of the fundamental timescales of star forming clouds introduced in Sec.~\ref{subsubsec:timescales},  the lag time, $t_{\rm lag}$, between when clouds begin to accumulate and when star formation begins. Taken at face value, they do not explain cloud lifetimes, $t_{\rm life}$, nor the low overall star formation rates or equivalently the long gas depletion timescales, $t_{\rm dep}$.  This is because these models by themselves lack an exit strategy. In the absence of energy sources within the cloud, the  accumulated mass will start to collapse globally, the clouds would settle and convert a substantial fraction of their mass into stars \citep{vazquez07,Heitsch08a}, violating not only the observed cloud lifetime limits, but also the observed limits on the star formation rate. Additional processes, most likely stellar feedback, are required to set the two remaining timescales. We come back to this issue in Section \ref{sec:feedback}.

\subsection{Modeling Molecular Cloud Formation in Hydrodynamic Simulations with Time-Dependent Chemistry}
The chemical composition of the ISM is complex. Over 120 different molecular species have been detected in interstellar space \citep{langer00} and while many of these are found in detectable amounts only in dense, well-shielded gas, there remain a significant number that have been detected in diffuse, unshielded gas \citep{oneill02}. A full chemical model of the ISM can easily involve several  hundred different atomic and molecular species and isotopologues and several thousand different reactions, even if reactions on grain surfaces are neglected, see e.g.\ the UMIST database \citep{leteuff00}.

\begin{figure*}[htbp]
\begin{center}
\includegraphics[width=0.4\textwidth]{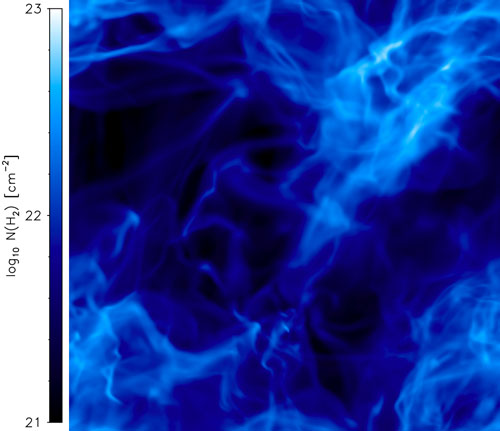}
\includegraphics[width=0.4\textwidth]{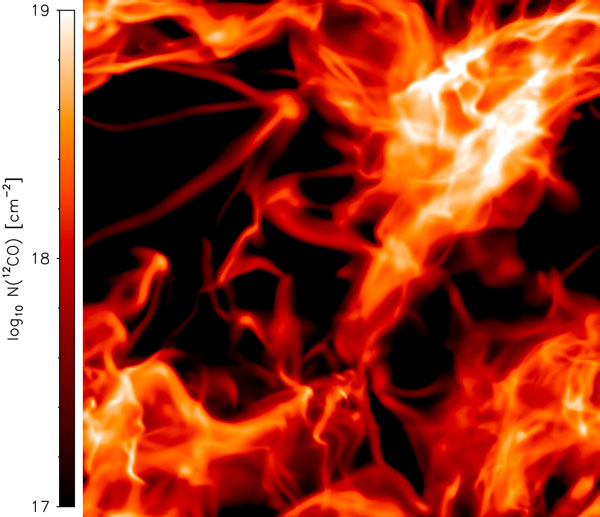}\\[0.5cm]
\includegraphics[width=0.4\textwidth]{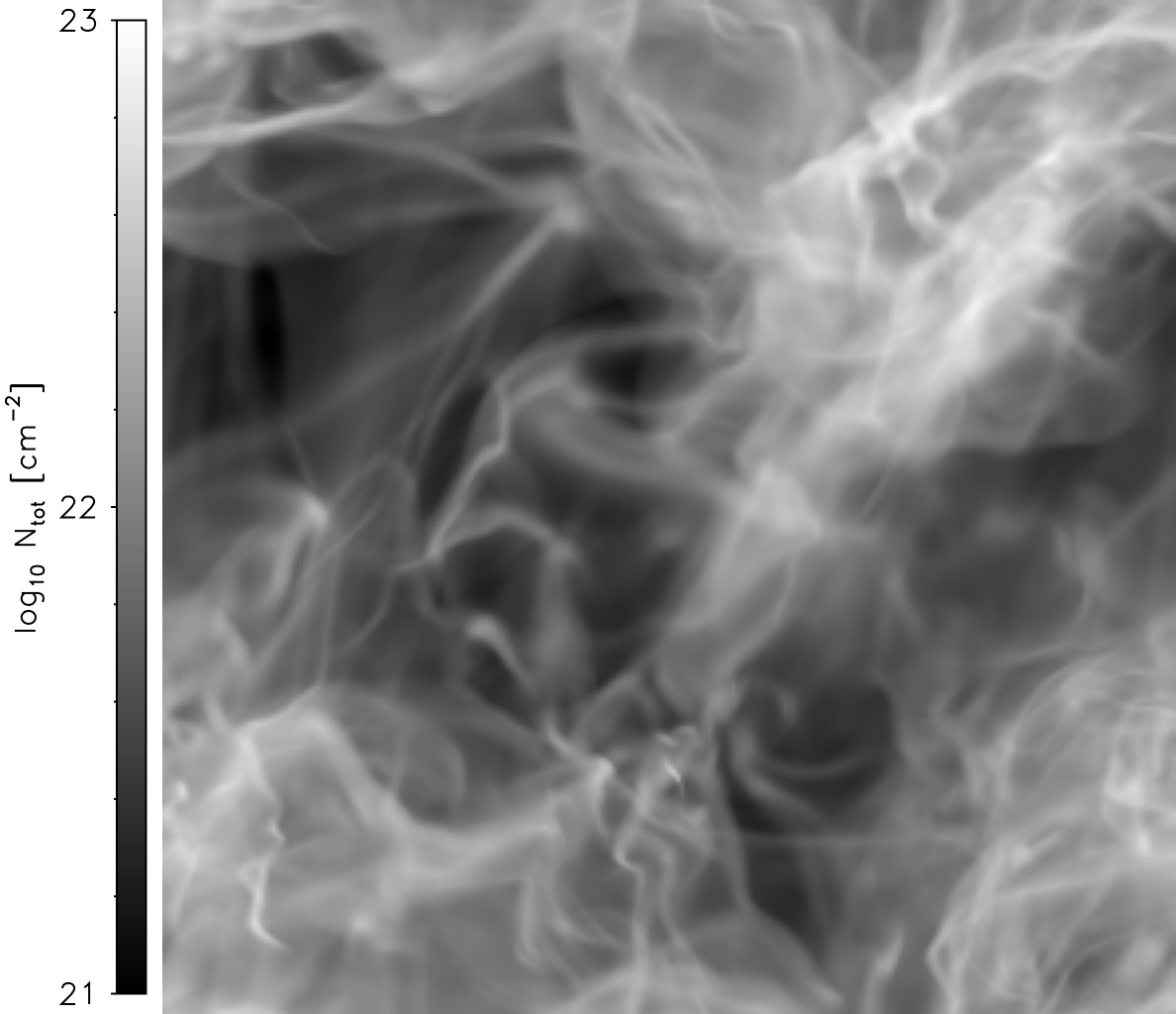}
\includegraphics[width=0.4\textwidth]{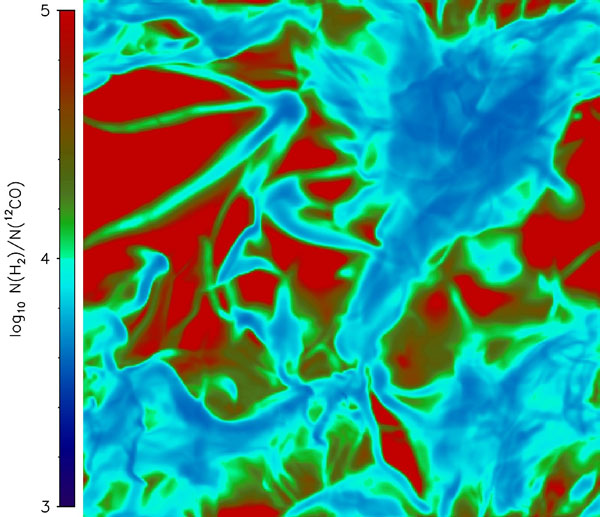}
\caption{Synthetic emission maps from turbulent-box calculations with time-dependent chemistry. The top left panel depicts the total H$_2$ column density, while the top right panel shows the integrated optically thin line emission from the tracer molecule CO. For comparison, the lower left image shows the total gas column density. Inspection of the top two images illustrates that CO traces the H$_2$ distribution very well in high column density regions, however, fails to do so for the low-density surface layers that are more strongly exposed to the external radiation field. This is quantified in the bottom right image which shows the ratio between both values. This ratio is related to the so-called ``x-factor'' that is commonly used to convert CO intensity maps into H$_2$ maps \citep{PinedaCaselliGoodman2008}. {\em (Image courtesy of C.\ Federrath)}}
\label{fig:column-densities}
\end{center}
\end{figure*}

It is currently impractical to incorporate this amount of chemistry into a 3D hydrodynamical code. The key to constructing time-dependent chemical networks that can run alongside the dynamic  evolution of the system therefore is to select a number of chemical species and mutual reaction rates that is small enough so that the chemical network can be solved in a short enough time so that it is tractable to do so during each system timestep and that is large and complete enough so that the overall evolution of the system is still described adequately. In the context of molecular cloud formation it is clearly necessary to be able to follow the formation and destruction of H$_2$ with a reasonable degree of accuracy. Beyond this, the only chemistry that is really required is that which plays a role in determining the thermal balance of the gas. In other words, we need only follow the chemistry of H$_2$, and of a few other major coolants such as C$^+$ or O in low 
column density gas, or CO in high column density regions. As few as thirty species and two hundred reactions appear to be sufficient for accurately modelling the most important hydrogen, carbon and oxygen chemistry in molecular cloud formation calculations \citep{glover09}, and a network of this size has been shown to be practical to incorporate into a 3D hydrodynamical code \cite{Jappsenetal07}.

Many reaction rates are sensitive to the external radiation field. Molecular hydrogen, for example, can only remain stable in regions where the column density is high enough to significantly attenuate the Galactic radiation field \citep{draine96}. This means, that any sensible calculation of chemical reaction rates requires knowledge of the local radiation field. Ideally, simulations with time-dependent chemistry running alongside the hydrodynamics should also include full radiative transfer (as discussed in Section \ref{subsec:modeling.feedback}). This is, however, beyond the capabilities of current numerical schemes, and most astrophysical applications treat radiation in a very approximate fashion only, for example, by assuming a constant background field or by computing column densities and optical depths only along the principle axes of the system. 

Although it has as yet received only limited attention from computational astrophysicists, efficient coupling between chemical reaction networks and hydrodynamic solvers is an active area of research in fields such as combustion modelling \citep[e.g.][]{rp07} or atmospheric chemistry \citep[e.g.][]{sportisse07}. The basic principles are straightforward. 
One usually uses some form of operator splitting to separate the treatment of the chemistry from the advection and/or diffusion terms. During the chemistry sub-step, one updates the chemical  abundances by solving a coupled set of rate  equations of the form 
\begin{equation}
\frac{{\rm d}n_{i}}{{\rm d} t} = C_{i} - D_{i} n_{i},
\end{equation}
where $n_{i}$ is the number density of species $i$, and $C_{i}$ and $D_{i}$ are chemical creation and destruction terms that generally depend on the temperature $T$ and the chemical abundances of the other reactants in the system. Most chemical reaction networks are stiff -- that is, they contain a wide range of different characteristic timescales -- and so to ensure stability, it is usually necessary to solve these coupled rate equations with an implicit scheme.
The simplest implicit techniques have a computational cost that scales as the cube of the number of species, and so considerable ingenuity has been expended in attempts to reduce this cost, for instance by making use of chemical conservation laws to reduce the number of species that need to be tracked, or by taking advantage of the typically sparse nature of the Jacobian of the coupled set of equations \citep[e.g.][]{nejad05}.

The thermal evolution of the gas is usually modelled using a library of cooling functions for each considered species. For example, state-of-the-art calculations include the effects of atomic fine structure cooling (e.g. C, C$^+$, O, Si and Si$^+$), rotational and vibrational cooling of the considered molecules (e.g.\ H$_2$, HD, CO and H$_2$O), Lyman-$\alpha$ cooling, Compton cooling, and H$^+$ recombination cooling, as well as other processes of lesser importance. The numerical implementation usually adopts the isochoric approximation \cite{springel01a} and computes emission strength using the large-velocity gradient approach, which assumes that the emitted lines are absorbed locally only. During a given hydrodynamic timestep one computes first $\dot{u}_{\rm ad}$, the rate of change of the internal energy due to adiabatic gas physics. Then one has to solve an implicit equation for the new internal energy of the form 
\begin{equation}
 u^{\rm new} = u^{\rm old} + \left[ \dot{u}_{\rm ad}  - \frac{\Lambda \left(\rho^{\rm new}, u^{\rm new}\right)}{\rho^{\rm new}}\right] \Delta t\,,
\end{equation}
where $u^{\rm new}$ and $u^{\rm old}$ are the internal energy per unit mass at the current and old time, respectively, $\rho^{\rm new}$ is the gas density at the current time. It is often necessary to solve this implicit equation simultaneously with the chemical rate equations.

\section{From Cloud Cores to Stars}
\label{sec:core-star}

\subsection{Observational Properties of Molecular Cloud Cores}
\label{subsec:obs}

\subsubsection{Statistical Properties}
\label{subsubsec:statistics}

Emission line observations and dust extinction maps of molecular clouds reveal extremely complex morphological structure with clumps and filaments on all scales accessible by present day telescopes. Typical parameters of different regions in molecular clouds are listed in Table \ref{tab:MC-prop}. The volume filling factor $n / \langle n \rangle$ (where $n$ is the local density, while $\langle n \rangle$ is the average density of the cloud) of dense clumps, even denser subclumps and so on, is rather small, ranging from 10\% down to 0.1\% at densities of $n >10^5$~cm$^{-3}$ \cite{blitz93,mckee99,williams00,beuther00a}.   This hierarchical configuration is often interpreted as being fractal \citep{ElmegreenFalgarone1996,StutzkiEtAl1998,bensch01a} which however, is still subject to debate \citep{blitz97}. It is important to note that star formation always occurs in the densest regions within a cloud, so only a small fraction of molecular cloud matter is actually involved in building up stars, while the bulk of the material remains at lower densities. This is most likely the key to explaining the low star formation efficiencies as discussed above in Section \ref{subsec:cloud-form}.

The mass spectrum of clumps in molecular clouds appears to be well described by a power law,
\begin{equation}
\label{eqn:clump-spectrum}
\frac{dN}{dm} \propto m^{\alpha}\:,
\end{equation}
with the exponent being in the range $-1.3 < \alpha < -1.8$, indicating self-similarity \citep{StutzkiGuesten1990,Williamsetal1994,kramer98a}. There is no natural mass or size scale between the lower and upper limits of the observations. The smallest observed structures are protostellar cores with masses of a few solar masses or less and sizes of $\sil 0.1\,$pc. The fact that all studies obtain a similar power law is remarkable, and may be the result of turbulent motions and thermal instability acting on self-gravitating gas.  Given the uncertainties in determining the slope, it appears reasonable to conclude that there is a universal mass spectrum for the clumps within a molecular cloud, and that the distribution is a power law within a mass range of three orders of magnitude, i.e.\ from $1\,$M$_{\odot}$ to about $1000\,$M$_{\odot}$. Hence, it appears plausible that the physical processes that determine the distribution of clump masses are rather similar from cloud to cloud. And vice versa, clouds that show significant deviation from this universal distribution most likely had different dynamical histories.

Most of the objects that enter in the above morphological analyses are not gravitationally bound \citep{StutzkiGuesten1990,Morata2005,klessen05,Dib2007}. It is interesting to note that the distribution changes as one probes smaller and smaller scales and more and more bound objects. When considering prestellar cores, which are thought to be the direct progenitors of individual stars or small multiple systems, then the mass function is well described by a double power law fit $dN/dm \propto m^{-\alpha}$ following $\alpha = 2.5$ above $\sim0.5\ $M$_{\odot}$ and  $\alpha = 1.5$ below.  The first large study of this kind was published by \citet{motte98}, for a population of submillimetre cores in $\rho$ Oph. Using data obtained with IRAM, they discovered a total of 58 starless clumps, ranging in mass from $0.05\ $M$_{\odot}$ to $\sim 3\ $M$_{\odot}$. Similar results are obtained from the Serpens cloud \citep{testi98}, for  Orion B North  \citep{johnstone01} and Orion B South \citep{Johnstoneetal2006}, or for the Pipe Nebula  \cite{Ladaetal2006}. Currently all observational data \cite{motte98,testi98,johnstone00,johnstone01,Johnstoneetal2006,NutterWardThompson2006,alves07, DiFrancescoetal2007,WardThompsonetal2007,lada08a}  reveal that the mass function of prestellar cores is strikingly similar in shape to the stellar initial mass function, the IMF. To reach complete overlap one is required to introduce a mass scaling or efficiency factor in the range 2 to 10, which differs in different regions. An exciting interpretation of these observations is that we are witnessing the direct formation of the IMF via fragmentation of the parent cloud. However, we note that the observational data also indicate that a considerable fraction of the prestellar cores do not exceed the critical mass for gravitational collapse, much like the clumps on larger scale. The evidence for a one-to-one mapping between prestellar cores and the stellar mass thus is by no means conclusive as we will discuss in more detail in Section \ref{subsec:IMF}.

\subsubsection{Individual Cores}
\label{subsubsec:ind.cores}

\textit{Density Structure.} 
The density structure of prestellar cores is typically estimated through the analysis of dust emission or absorption using near-IR extinction mapping of background starlight, mapping of millimeter/submillimeter dust continuum emission, and  mapping of dust absorption against the bright mid-IR background emission \cite{bergin07}. A main characteristic of the density profiles derived with the above techniques is that they require a central flattening. The density gradient of a core is flatter than r$^{-1}$ within radii smaller than 2500 -- 5000 AU, and that the typical central density of a core is 10$^{5}$ -- 10$^{6}$ cm$^{-3}$ \cite{motte98,ward-thompson99}.  A popular approach is to describe these cores as truncated isothermal (Bonnor-Ebert) sphere \cite{ebert55,bonnor56}, that often (but not always) provides a good fit to the data \cite{bacman01,Alves01,kandori05}. These are equilibrium solutions of self-gravitating gas spheres bounded by external pressure. However, such density structure is not unique. Numerical calculations of the dynamical evolution of supersonically turbulent clouds show that transient cores forming at the stagnation points of convergent flows exhibit similar morphology \cite{ballesteros03}.

\begin{figure*}[htbp]
\begin{center}
\includegraphics[height=4.5cm]{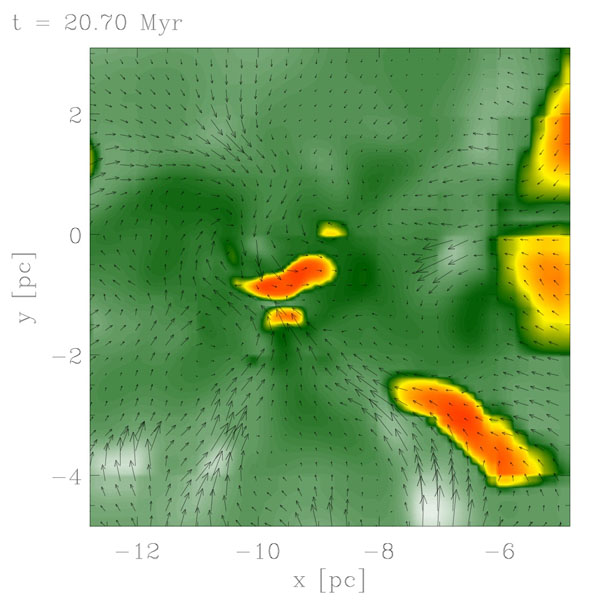}
\includegraphics[height=4.5cm]{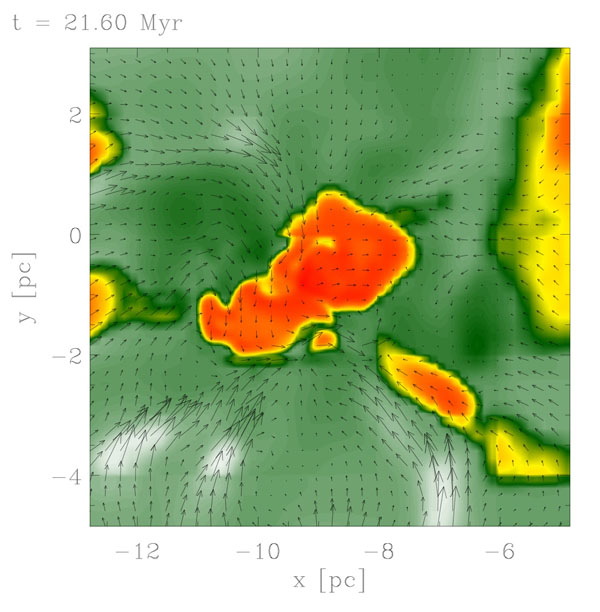}
\includegraphics[height=4.5cm]{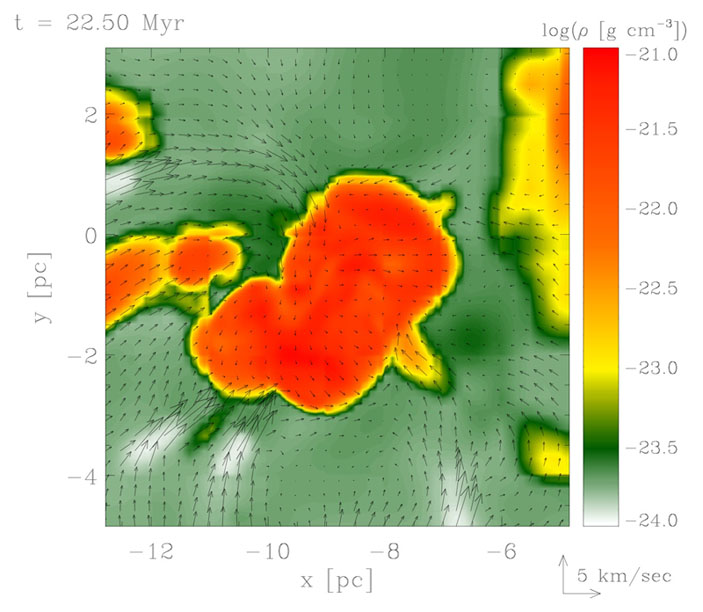}
\caption{Formation and growth of molecular cloud cores by thermal instability triggered by a large-scale convergent flow: A small cold condensate grows from the thermally unstable warm neutral medium by outward propagation of its boundary layer. Coalescence and merging with nearby clumps further increases its mass and size. The global gravitational potential of the proto-cloud enhances the merging probability with time. The images show 2D slices of the density (logarithmic colour scale) and the gas velocity (indicated as arrows) in the plane perpendicular to the large scale flow. {\em (From a numerical simulation by \citet{Banerjee08b})}}
\label{fig:clumps}
\end{center}
\end{figure*}

\textit{Thermal Stucture.}
The kinetic temperature of the dust and gas components in a core is  regulated by interplay between various heating and cooling processes. At high densities ($>10^5$ cm$^{-3}$) in the inner part of the cores, the gas and dust have to be coupled thermally via collisions \cite{goldsmith78,burke83,Goldsmith01}. At lower densities, which correspond to the outer parts of the cores, the two temperatures are not necessary expected to be the same. Thus, the dust and gas temperature distributions need to be inferred from observations independently. Large-scale studies of dust temperature show that the grains in starless cores are colder than in the surrounding lower-density medium. Far-IR observations toward the vicinity of a number of dense cores provide evidence for flat or decreasing temperature gradients with cloud temperatures of $15 -20\,$K and core values of $8 -12\,$K \cite{Ward02,toth04}. These observations are consistent with dust radiative transfer modeling in cores illuminated by interstellar radiation field \cite{langer05,keto05,stamatellos07a}, where the dust temperature is $\sim 7\,$K in the core center and increases up to 16$\,$K in the envelope.  The gas temperature in molecular clouds and cores is commonly infered from the level excitation of simple molecules like CO and NH$_3$ \cite{evans99,walmsley83}. One finds gas temperatures of 10--15 K, with a possible increase toward the lower density gas near the cloud edges. It is believed that the gas heating in prestellar cores mostly occurs through ionization by cosmic rays, while the cooling is mainly due to line radiation from molecules, especially CO \cite{goldsmith78}. 
Altogether, the fact that prestellar cores are cold and roughly isothermal with at most a modest increase in temperature from the center to the edge is consistent with numerical models of cores forming from thermal instability \cite{heitsch06a,keto08,banerjee08}, see also Figure \ref{fig:clumps}. 

\begin{figure*}[htbp]
\begin{center}
\includegraphics[width=0.9\textwidth]{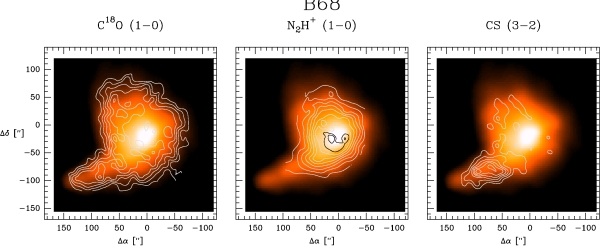}
\caption{Maps of molecular line emission from  C$^{18}$O, N$_2$H$^+$, and CS superimposed on a dust extinction maps of the dark cloud core Barnard 68 \cite{Alves01,bergin02,lada03b}. The three images illustrate the effects of depletion onto grains in the high-density central region of the core. N$_2$H$^+$ is the least and CS the most depleted species. {\em (Image courtesy of E.\ A.\ Bergin)}
}
\label{fig:B68}
\end{center}
\end{figure*}

\textit{Chemical Stucture.}
Maps of integral line intensity  can look very different for different molecular tracers. In particular, the N$_2$H$^+$ and NH$_3$ emission more closely follows the dust emission while the C$^{18}$O and CS emission appears as a ``ring-like''\, structure around the dust emission maximum \cite{bergin02,tafalla02,lada03b,maret07a}. For illustration see Figure \ref{fig:B68}. The common theoretical interpretion of these data is that carbon-bearing species, represented by CO and CS freeze-out on the dust grains from the gas while the abundances of nitrogen-hydrogen bearing molecules, N$_2$H$^+$ and NH$_3$, either stay constant or decay more slowly.  At the same time, chemical models of prestellar cores predict that molecules in the core envelope have to be destroyed by interstellar UV field \cite{pavlyuchenkov06,aikawa08}. The chemical stratification significantly complicates the interpretation of molecular line observations and again requires the use of sophisticated chemical models which have to be 
coupled to the dynamical evolution. From observational side, the freeze-out of many molecules makes it difficult to use their emission lines for probing the physical conditions in the inner regions of the cores.  At the same time, the modeling of the chemical evolution can provide us with the important parameters of the cores. For example, the level of CS depletion can be used to constrain the age of the prestellar cores while the deficit of CS in the envelope  can indicate the strength of the external UV field \citep{bergin07}. In any case, any physical interpretation of the molecular lines in prestellar cores has to be based on chemical models and should do justice to the underlying density and velocity pattern of the gas.

\textit{Kinematic Stucture.}
In contrast to the supersonic velocity fields observed in molecular clouds, dense cores have low internal velocities. Starless cores in clouds like Taurus, Perseus, and Ophiuchus systematically present spectra with close-to-thermal linewidths, even when observed at low angular resolution \cite{myers83,jijina99}. This indicates that the gas motions inside the cores are subsonic or at best transsonic, i.e.\ with Mach numbers $\sil 2$  \cite{kirk07, andre07, rosolowsky08a}. 
In some cores also inward motions have been detected. They are inferred from the observation of optically thick, self-absorbed lines of species like CS, H$_2$CO, or HCO$^+$, in which low-excitation foreground gas absorbs part of the background emission. Typical inflow velocities are of order of $0.05-0.1\,$km/s  and are observed on scales of  $0.05 - 0.15\,$pc, comparable to the observed size of the cores \cite{lee99}. The overall velocity structure of starless cores appears broadly consistent with the structure predicted by models in which protostellar cores form as the stagnation points of convergent flows, but the agreement is not perfect. Simulations of core formation do correctly find that most cores are at most transsonic \cite{klessen05, offner08b}, but the distribution of velocity dispersions has a small tail of highly supersonic cores that is not observed. Clearly more theoretical and numerical work is needed. In particular, the comparison should be based on synthetic line emission maps, which requires to couple a chemical network and radiative transfer to the simulated density profiles as discussed above. In addition, it is also plausible that the discrepancy occurs because the simulations do not include all the necessary physics such as radiative feedback and magnetic fields. Subsonic turbulence contributes less to the energy budget of the cloud than thermal pressure and so cannot provide sufficient support against gravitational collapse \cite{myers83,goodman98,tafalla06}. If cores are longer lasting entities there must be other mechanisms to provide stability. Obvious candidates are magnetic fields \cite{shu87}. However, they are usually not strong enough to provide sufficient support \cite{crutcher99a,crutcher00,bourke01,crutcher08} as discussed below. Most observed cores are thus likely to be evolving transient  objects that never reach any equilibrium state.  

\textit{Magnetic Field Structure.}
Magnetic fields are ubiquitously observed in the interstellar gas on all scales \citep{Crutcher03,heiles05}. However, their importance for star formation and for the morphology and evolution of molecular cloud cores remains controversial. A crucial parameter in this debate is the ratio between core mass and magnetic flux. 
In supercritical cores, this ratio exceeds a critical value and collapse can proceed. In subcritical ones, magnetic fields provide  stability \citep{spitzer78,mouschovias91b,mouschovias91a}. 
Measurements of the Zeeman splitting of molecular lines in nearby cloud cores indicate mass-to-flux ratios that lie above the critical value, in some cases only by a small margin but very often by factors of many  if non-detections are included \citep{crutcher99, bourke01, crutcher08}. The polarization of dust emission offers an alternative pathway to studying the magnetic field structure of molecular cloud cores. MHD simulations of turbulent clouds predict degrees of polarization between 1 and 10\%, regardless of whether turbulent energy dominates over the magnetic energy (i.e.\ the turbulence is super-Alfv\'enic) or not \citep{padoan99,padoan01}. However, converting polarization into magnetic field strength is very difficult \citep{heitsch01b}.  Altogether, the current observational finding imply that magnetic fields must be considered when studying stellar birth, but also that they are not the dominant agent that determines when and where stars form within a cloud. Magnetic fields appear too weak to prevent gravitational collapse to occur.

This conclusion means that in many cases and to reasonable approximation purely hydrodynamic calculations are sufficient for star formation simulations. However, when more precise and quantitative predictions are desired, e.g.\ when attempting to predict star formation timescales or binary properties, it is necessary to perform magnetohydrodynamic (MHD) simulations or even consider non-ideal MHD. The latter means to take ambipolar diffusion (drift between charged and neutral particles) or Ohmic dissipation into account. Recent numerical simulations have shown that even a weak magnetic field can have noticeable dynamical effects. It can alter how cores fragment \cite{price07a, price08a, hennebelle08a, hennebelle08c}, change the coupling between stellar feedback processes and their parent clouds \cite{nakamura07, krumholz07f}, influence the properties of protostellar disks due to magnetic braking \cite{price07b, mellon08a, mellon08b}, or slow down the overall evolution \cite{Heitschetal2001}.

\subsubsection{Models of Cloud Evolution and Star Formation}
There are two main competing models that describe the evolution of the cloud cores.  It was proposed in the 1980's that cores in low-mass star-forming regions evolve quasi-statically in magnetically subcritical clouds \cite{shu87}. Gravitational contraction is mediated by ambipolar diffusion \cite{mouschovias76,mouschovias79,mouschovias81,mouschovias91b} causing a redistribution of magnetic flux until the inner regions of the core become supercritical and go into dynamical collapse.  This process was originally thought to be slow, because in highly subcritical clouds the ambipolar diffusion timescale is about 10 times larger than the dynamical one. However for cores close to the critical value, as is suggested by observations, both timescales are comparable. Numerical simulations furthermore indicate that the ambipolar diffusion timescale becomes significantly shorter for turbulent velocities similar to the values observed in nearby star-forming region  \citep{FatuzzoAdams2002,heitsch04,zli04}. The fact that ambipolar diffusion may not be a slow process under realistic cloud conditions, as well as the fact that most cloud cores are magnetically supercritical \cite{crutcher99a,crutcher00,bourke01,crutcher08}  has cast significant doubts on any magnetically-dominated quasi-static models of stellar birth.

For this reason, star-formation research has turned into considering supersonic turbulence as being on of the principal physical agent regulating stellar birth. The presence of turbulence, in particular of supersonic turbulence, has important consequences for molecular cloud evolution. On large scales it can support clouds against contraction, while on small scales it can provoke localized collapse. Turbulence establishes a complex network of interacting shocks, where dense cores form at the stagnation points of convergent flows. The density can be large enough for gravitational collapse to set in. However, the fluctuations in turbulent velocity fields are highly transient.  The random flow that creates local density enhancements can disperse them again.  For local collapse to actually result in the formation of stars, high density fluctuations must collapse on timescales shorter than the typical time interval between two successive shock passages.  Only then are they able to `decouple' from the ambient flow and survive subsequent shock interactions.  The shorter the time between shock passages, the less likely these fluctuations are to survive. Hence, the timescale and efficiency of protostellar core formation depend strongly on the wavelength and strength of the driving source \citep{klessen00b,Heitschetal2001,Vazquez03,maclow04,krumholz05c,ballesteros07b,mckee07}, and accretion histories of individual protostars are strongly time varying \citep{Klessen2001b,SchmejaKlessen2004}.

Interstellar turbulence is observed to be dominated by large-scale modes \citep{maclow00b,ossenkopf01a,Ossenkopf02}. This implies it is very efficient  in sweeping up molecular cloud material, thus creating massive coherent structures. The result is a large region in which many Jeans masses of material become unstable to collapse at about the same time, leading to coherent structures in the forming stars. This is a likely explanation for the observed clustering of young stars \cite{klessen00b}, as we discuss in the following section.

\subsection{Spatial Distribution}
\label{subsec:spatial-dist}

The advent of sensitive infrared detectors in the last decade has made it possible to perform wide-area surveys. These have led us to recognize that most stars form in clusters and aggregates of various size and mass scales, and that isolated or widely distributed star formation is the exception rather than the rule \citep{lada03}. The complex hierarchical structure of molecular clouds (Figure \ref{fig:perseus}) provides a natural explanation for this finding. 
Star-forming molecular cloud cores vary enormously in size and mass. In small, low-density, clouds stars form with low efficiency, more or less in isolation or scattered around in small groups of up to a few dozen members. Denser and more massive clouds may build up stars in associations and clusters of a few hundred members.  This appears to be the most common mode of star formation in
the solar neighborhood \citep{adams01}. Examples of star formation in small groups and associations are found in the Taurus-Aurigae molecular cloud \citep{hartmann02}. Young stellar groups with a few hundred members form in the Chamaeleon I \citep{persi00} or $\rho$-Ophiuchi \citep{Bontempsetal2001} dark clouds. Each of these clouds is at a distance of about 130 to 160$\,$pc from the Sun.  Like most of the nearby young star forming regions they appear to be associated with a ring-like structure in the Galactic disk called Gould's belt \citep{poeppel97}.

\begin{figure}
\includegraphics[width=\columnwidth]{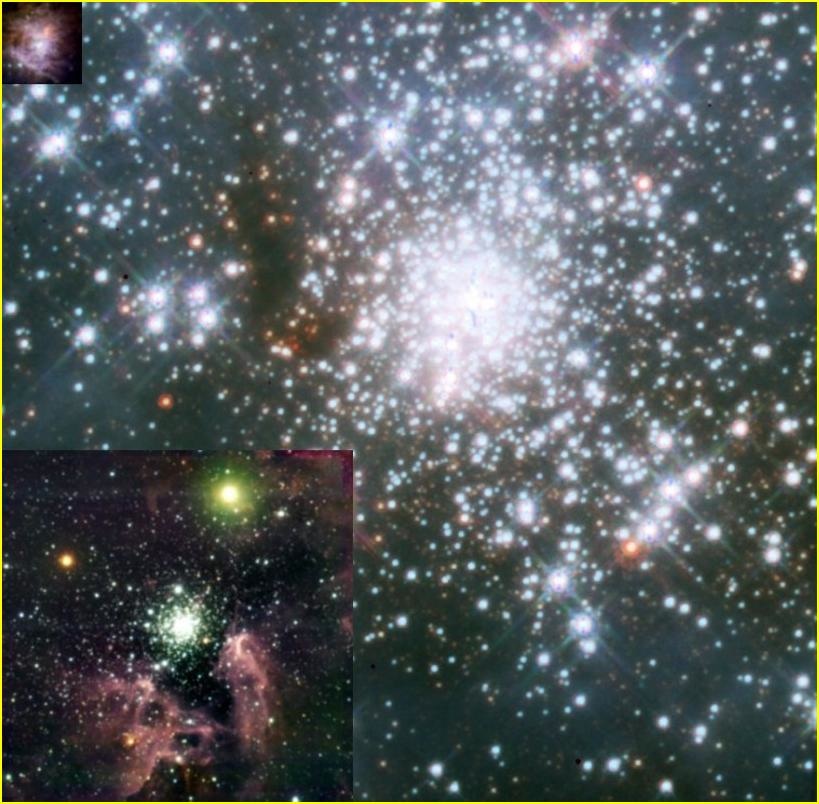}
\caption{\label{fig:clusters}
Comparison of clusters of different masses scaled to same relative distance. 
The cluster in the upper left corner is the Orion Nebula Cluster \citep{mccaughrean01} and the one at the lower left is NGC~3603 \citep{brandl99}, both observed with the Very Large Telescope at infrared wavelength. The large cluster in the center is 30 Doradus in the LMC observed with the Hubble Space Telescope (courtesy of M.\ J.\ McCaughrean). The total mass increases roughly by a factor of ten from one cluster to the other. {\em (Composite image courtesy of H. Zinnecker)}
}
\end{figure}

The formation of dense rich clusters with thousands of stars is rare. The closest region where this happens  is the Orion Nebula Cluster in L1641 \citep{Hillenbrand97,hillenbrand98}, which lies at a distance of $410\,$pc
\cite{sandstrom07, menten07, hirota07, caballero08}.  A rich cluster somewhat further away is associated with the Monoceros R2 cloud \citep{carpenter97} at a distance of $\sim 830\,$pc.  The cluster NGC~3603 is roughly ten times more massive than the Orion Nebula Cluster.  It lies in the Carina region, at about $7\,$kpc distance. It contains about a dozen O stars, and is the nearest object analogous to a starburst knot \citep{brandl99,moffat02}. To find star-forming regions building up hundreds of O stars one has to look towards giant extragalactic H\textsc{ii} regions, the nearest of which is 30 Doradus in the Large Magellanic Cloud, a satellite galaxy of our Milky Way at a distance at 55$\,$kpc. The giant star forming region 30 Doradus is thought to contain up to a hundred thousand young stars, including more than 400 O stars \citep{hunter95,walborn97,townsley06}. 
This sequence as depicted in Figure \ref{fig:clusters} demonstrates that the star formation process spans many orders of magnitude in scale, ranging from isolated single stars to massive young clusters with several $10^4$ stars.

\subsection{The Stellar Initial Mass Function and other Statistical Characteristics of Star Formation}
\label{subsec:IMF}

The mass distribution of young stars follows a well-known distribution called the Initial Mass Function (IMF). For stellar masses $m \ge 1 M_{\odot}$ it shows a power-law behavior $dN/dm \propto m^{\alpha}$ , with slope $\alpha = -2.3$ \citep{Salpeter1955,scalo98,kroupa02,chabrier03}. Below $1 M_{\odot}$, the 
IMF flattens, a change in behavior than can be represented either as a lognormal \cite{chabrier03} or a change in power law index \cite{kroupa02}. 
At the extreme ends of the stellar mass spectrum, however, our knowledge of both the IMF are limited. Massive stars are very rare and rather short lived. The number of massive stars that are sufficiently near to study in detail and with very high spatial resolution, for example to determine  multiplicity, therefore is small  \citep{zinnecker07}. Low-mass stars and brown dwarfs, on the other hand, are faint, so they too are difficult to study in detail \citep{burrows01}. Such studies, however, are in great demand, because secondary indicators such as the fraction of binaries and higher-order multiples as function of mass, or the distribution disks around very young stars or possible signatures of accretion during their formation are probably better suited to distinguish between different star-formation models than just looking at the IMF. 
In contrast to the observational agreement on the IMF, at least above the substellar regime, there is still considerable disagreement on the theoretical side. The origin of the IMF is a major topic of theoretical research which we examine only briefly here to give the necessary theoretical background for our discussion of numerical work. 
Other reviews provide considerably more detail \citep{maclow04,bonnell07a,Larson07,mckee07}.

Early models for the origin of the IMF generally relied on statistical arguments, appealing to random process of collapse in a fractal cloud \citep{Elmegreen1997,elmegreen00b}, or to the central limit theorem to explain its characteristic shape \citep{lars73,zinnecker84}. Researchers have also invoked feedback processes that cut off accretion onto individual protostars \cite{AdamsFatuzzo1996}. Today, however, there are three dominant schools of thought regarding the origin on the IMF, although the boundaries between these pictures are not clearly defined, and numerous hybrid models have been proposed.

One model, called core accretion, takes as its starting point the striking similarity between the shape of the observed core mass distribution and the IMF. This model posits that there is a one-to-one relation between the distributions, so that individual cores are the progenitors of individual stars or star systems. The factor of $\sim\,$3 decrease in mass between cores and stars is the result of feedback processes, mostly protostellar outflows, that eject a fixed fraction of the mass in a core rather than letting it accrete onto the star \cite{matzner00}. This model reduces the problem of the origin of the IMF to the problem of determining the mass spectrum of bound cores, although strictly speaking the idea that the IMF is set by the mass spectrum of cores is independent of any particular model for the origin of that mass spectrum. Arguments to explain the core mass distribution generally rely on the statistical properties of turbulence \cite{Klessen2001, Padoan02, HennebelleChabrier2008}, which generate structures with a pure powerlaw mass spectrum. The thermal Jeans mass in the cloud then imposes the flattening and turn-down in the observed mass spectrum.

A second model for the origin of the IMF, called competitive accretion, focuses instead in interaction between protostars, and between a protostellar population and the gas cloud around it \cite{bonnell01a, bonnell01b, bonnell02, bate03, bate05, bonnell06d, bonnell08a}. In the competitive accretion picture the origin of the peak in the IMF is much the same as in the core accretion model: it is set by the Jeans mass in the prestellar gas cloud. However, rather than fragmentation in the gas phase producing a spectrum of core masses, each of which collapses down to a single star or star system, in the competitive accretion model all gas fragments down to roughly the Jeans mass. Prompt fragmentation therefore creates a mass function that lacks the powerlaw tail at high masses that we observe in the stellar mass function. This part of the distribution forms via a second phase in which Jeans mass-protostars compete for gas in the center of a dense cluster. The cluster potential channels mass toward the center, so stars that remain in the center grow to large masses, while those that are ejected from the cluster center by $N$-body interactions remain low mass \cite{klessen00a,bonnell04}. In this model, the apparent similarity between the core and stellar mass functions is an illusion, because the observed cores do not correspond to gravitationally bound structures that will collapse to stars \cite{clark06, smith08a}.

One potential drawback to both the core accretion and competitive accretion models is that they rely on the Jeans mass to determine the peak of the IMF, but leave unanswered the question of how to compute it. This question is subtle because molecular clouds are nearly isothermal, but they contain a very wide range of densities, and it is unclear which density should be used. A promising idea to resolve this question, which is the basis for a third model of the IMF, focuses on the thermodynamic properties of the gas.  The amount of fragmentation occurring during gravitational collapse depends on the compressibility of the gas \citep{li03}.  For polytropic indices $\gamma < 1$, turbulent compressions cause large density enhancements in which the   Jeans mass falls substantially, allowing many fragments to collapse. Only a few massive fragments get compressed strongly enough to collapse in less compressible gas though. In real molecular gas, the   compressibility varies as the opacity and radiative heating  increase. \citet{Larson05} noted that the thermal coupling of the gas  to the dust at densities above $n_{\rm crit} \sim 10^5-10^6$~cm$^{-3}$  leads to a shift from an adiabatic index of $\gamma \sim 0.7$ to 1.1 as the density increases above $n_{crit}$.  The Jeans mass evaluated at the temperature and density where this shift occurs sets a mass scale for the peak of the IMF. The apparent universality of the IMF in the Milky Way and nearby galaxies may be caused by the insensitivity of the dust temperature on the intensity of the interstellar radiation field \citep{elmegreen08}.  Not only does this mechanism set the peak mass, but also appears to produce a power-law distribution of masses at the high-mass end comparable to the observed distribution \citep{Jappsen05}.

Each of these models has potential problems. In the core accretion picture, hydrodynamic simulations seem to indicate that massive cores should fragment into many stars rather than collapsing monolithically \cite{dobbs05, clark06, bonnell06}. The hydrodynamic simulations almost certainly suffer from over-fragmentation because they do not include radiative feedback from embedded massive stars \cite{krumholz06b,krumholz07a,krumholz08a}, but no simulation to date has successfully formed a massive core in a turbulent cloud and followed it all the way to the formation of a massive star. 

In addition, the suggestion of a one-to-one mapping between the observed clumps and the final IMF is subject to strong debate. Many of the prestellar cores discussed in Section \ref{subsubsec:statistics} appear to be stable entities \citep{johnstone00,johnstone01,Johnstoneetal2006,lada08a}, and thus are unlikely to be in a state of active star formation. In addition, the simple interpretation that one core forms on average one star, and that all cores contain the same number of thermal Jeans masses, leads to a timescale problem \cite{clark07} that requires differences in the core mass function and the IMF. 

The criticism regarding neglect of radiative feedback effects also applies to the gas thermodynamic idea: the cooling curves which that model assumes in order to derive the Jeans mass ignore the influence of protostellar radiation on the temperature of the gas, which simulations show can suppress fragmentation in at least some circumstances \cite{krumholz07a}. The competitive accretion picture has also been challenged, on the grounds that the kinematic structure observed in star-forming regions is inconsistent with the idea that protostars have time to interact with one another strongly before they completely accrete their parent cores \cite{krumholz05e,andre07}.

\subsection{Modeling Cloud Fragmentation and Protostellar Collapse}
\label{subsec:models}

To adequately model the fragmentation of molecular clouds, the formation of dense cloud cores, the collapse of the gravitationally unstable subset of cores, and finally the build-up and mass growth of embedded protostars in their interior is an enormous computational challenge. It requires to follow the evolution of self-gravitating, highly turbulent gas over many order of magnitudes in density and lengthscale. Owing to the stochastic nature of supersonic turbulence, it is not know in advance where and when local collapse occurs. One therefore needs highly flexible numerical methods for solving the equations of hydrodynamics, schemes that can provide sufficient degrees of precision and resolution throughout the entire computational domain in an adaptive fashion. 

The star formation community is following two highly complementary approaches to reach these goals. One set of methods is based on dividing the computational domain into small volume elements and follow the fluxes of all relevant quantities from one cell to the other. These grid-based methods adopt an Eulerian point of view, because the flow is followed from fixed positions in space. A popular alternative is to split the model cloud into individual parcels of gas and follow their mutual interaction and evolution. Particle-methods therefore correspond to a Lagrangian point of view following the trajectories of individual fluid elements. 

\subsubsection{Grid-Based Methods}
\label{subsubsec:grid-methods}

The mathematical formulation of hydrodynamics consists of a set of partial differential equations that relate different flow properties (such as density and velocity) with each other and with thermodynamic quantities  (e.g.\ pressure, temperature or internal energy of the medium). They can be formulated in conservative form corresponding to the conservation of mass, momentum, and energy. As the number of variables is larger than the number of equations in the system, an additional equation is needed to find a unique solution. This closure relation is called equation of state and usually specifies the pressure as function of other thermodynamic variables \citep{LL-X}. In a broad sense, the hydrodynamics equations describe how signals propagate through a medium. They specify how local quantities relate to fluxes, e.g.\ how the density in some control volume depends on the mass flux through its surface. Equations of this kind are called hyperbolic equations.

Numerical solutions to partial differential equations always require discretization of the problem. This means that instead of continuous space and time dimensions we consider a discrete set of points. The computational domain is subdivided into individual volume elements surrounding node points on a grid or unstructured mesh. Finite volume methods are procedures for representing and evaluating partial differential equations as algebraic equations. They play a key role in computational fluid dynamics. Similar to finite difference schemes, values are calculated at discrete places on a meshed geometry. Volume integrals that contain a divergence term are converted to surface integrals, using the divergence theorem. These terms are then evaluated as fluxes at the surfaces of each volume element. Because the flux entering a given volume is identical to that leaving the adjacent volume, these methods are conservative. Finite volume methods have been in the focus of applied mathematics for decades. They have well defined convergence properties and available code packages have reached a very high degree of maturity and reliability. 

Finite volume schemes are most easily formulated for Cartesian grids with fixed cell size. The cell size determines the spatial resolution of the code. Wherever  higher resolution is needed, it can be achieved by refining the grid. We speak of adaptive mesh refinement (AMR), when this is done in an automated and locally adjustable way. There are a number of different approaches to AMR in the literature \citep{plewa05}. Most AMR treatments are based on finite-element models on unstructured meshes. They have the advantage to adapt easily to arbitrary complicated boundaries, however, constructing the mesh is very time consuming. When using Cartesian grids, one can refine on individual cells or on  larger groups of cells, so-called blocks \citep{berger84,berger89,bell94}. Cartesian AMR codes nowadays belong to the standard repertoire of numerical star formation studies.

In the following we list a few popular hydrodynamic and magnetohydrodynamic codes that have been developed in the past decade. All but two are freely available, although in some cases registration is needed before being able to download it from the web.  ZEUS-MP (http://cosmos.ucsd.edu/lca-www/software/index.html) is a parallel, non-adaptive hydro- and magnetohydrodynamics code with self-gravity and radiation \citep{norman00,hayes06}. NIRVANA (htpp://nirvana-code.aip.de) is an AMR code for non-relativistic, compressible, time-dependent, ideal or nonideal (viscosity, magnetic diffusion, thermal conduction) MHD \citep{ziegler05}. FLASH (http://flash.uchicago.edu/) is a highly modular, parallel adaptive-mesh code initially  designed for thermonuclear runaway problems but also capable of a wide variety of astrophysical problems \citep{FryxellEtAl2000}. 
ENZO (http://lca.ucsd.edu/projects/enzo) is a hybrid AMR code (hydrodynamics and $N$-body) which is designed to do simulations of cosmological structure formation \citep{ENZO04}. It has been extended to include magnetic fields, star formation, and ray-tracing radiation transfer. ATHENA (http://www.astro.princeton.edu/jstone/athena) is an MHD code built on a flexible framework that is designed to allow easy and modular extension to include a wide variety of physical processes \citep{stone08}.
The public version is non-adaptive, contains only hydrodynamics and MHD, and uses a fixed Cartesian grid, but extensions exist for non-Cartesian grids, for static and adaptive mesh refinement, for gravity, and for ionizing radiative transfer \cite{krumholz07f}.  Another widely used AMR code for MHD calculations is RAMSES \citep{teyssier02,fromang06}. It is very versatile with applications ranging from the the two-phase interstellar medium, to star and planet formation, as well as cosmological structure formation. A  grid code use commonly in star formation simulations, but which is not publicly available, is ORION: a parallel hydrodynamics code that includes self-gravity, sink particles coupled to a protostellar evolution model, and diffuse radiative transfer \cite{truelove98, klein99, fisher02, krumholz04, krumholz07b}. Last in our short summary is Proteus, a finite-volume method based on a gas-kinetic
formulation of the microscopic transport properties \citep{prendergastxu93,slyz1999,xu2001}. 
This approach allows the user to fully control the dissipative effects, making the scheme very attractive for e.g. turbulent 
transport studies. However, adding additional physics is generally more complicated than for other schemes. 
Proteus is fully parallelized and includes self-gravity, magnetic fields and a two-fluid model for ambipolar drift
\citep{heitsch04,heitsch07,Heitsch08a}.

\subsubsection{Particle-Based Methods}
\label{subsubsec:particle-methods}

Using a particle based scheme to solve the equations of hydrodynamics was first introduced by \citet{lucy77} and proposed independently by \citet{ging77}. Originally envisioned as a Monte-Carlo approach to calculate the time evolution of a hydrodynamic system, the formalism of \textit{smoothed particle hydrodynamics} (SPH)  is more intuitively understood as a particle interpolation scheme \citep{ging82}.  This provides better estimates for the errors involved and the convergence properties of the method. Excellent overviews of the method and some of its applications provide the reviews of \citet{benz90} and \citet{mona92, monaghan05}.

In the framework of classical physics, fluids and gases are large ensembles of interacting particles with the state of the system being described by the probability distribution function in phase space. Its time evolution is governed by Boltzmann's equation \citep{LL-X}. Hydrodynamic quantities can then be obtained in a local  averaging process involving scales larger than the local mean-free path. A related approach is facilitated in SPH. The fluid is represented by an ensemble of particles $i$, each carrying mass, momentum, and hydrodynamical properties.  The technique can therefore be seen as an extension to the well known $N$-body methods used in stellar dynamics.  Besides being characterized by its mass $m_i$ and velocity ${\mbox{\bf {\em v}}}_i$ and its location ${\mbox{\bf {\em r}}}_i$, each particle is associated with a density $\rho_i$, an internal energy $\epsilon_i$ (equivalent to a temperature $T_i$), and a pressure $p_i$. The time evolution of the fluid is then represented by the time evolution of the  SPH particles. Their behavior is governed by the equation of motion, supplemented by further equations to modify the hydrodynamical properties. Thermodynamical observables are obtained by averaging over an appropriate subset of the SPH particles.

Mathematically, the local averaging process for a quantity $f({\mbox{\bf {\em r}}}\:\!)$ can be performed by convolution with an appropriate  smoothing function  $W({\mbox{\bf {\em r}}},\mbox{{\bf {\em h}}}\:\!)$:
\begin{equation}
  \label{eqn:sph-1}
\left< f({\mbox{\bf {\em r}}}\:\!) \right> \equiv \int f({\mbox{\bf {\em r}}}\,') W({\mbox{\bf {\em r}}} -{\mbox{\bf {\em r}}}\,',{\mbox{\bf {\em h}}}\:\!)\,
d^3r'\:.   
\end{equation}
This function $W({\mbox{\bf {\em r}}},{\mbox{\bf {\em h}}}\:\!)$ is often referred to as the smoothing kernel. It must be normalized and approach the Dirac delta fuction in the limit  ${\mbox{\bf {\em h}}} \longrightarrow 0$.  For simplicity, most authors adopt spherical symmetry in the smoothing and averaging process, i.e.~the kernel degrades to an isotropic function of the interparticle distances: $W({\mbox{\bf {\em r}}},{\mbox{\bf {\em h}}}\:\!) \equiv W(r,h)$ with $r = |{\mbox{\bf {\em r}}}\,|$ and $h = |{\mbox{\bf {\em h}}}\:\!|$. 

The basic concept of SPH is a particle representation of the fluid. Hence, the spatial integration in the averaging process transforms into a summation over a fixed number of points. For example, the density
at the position of particle $i$ is computed as
\begin{equation}
\label{eqn:SPH-density}
\left< \rho({\mbox{\bf {\em r}}_i})\right> =   \sum_j m_j
W(|{\mbox{\bf {\em r}}_i} -{\mbox{\bf {\em r}}_j}|,h) \;.
\end{equation}
In this picture, the mass of each particle is smeared out over the kernel region. The continuous density distribution of the fluid is then obtained by summing over the local contribution from neighboring elements $j$. The name ``smoothed particle hydrodynamics'' derives from this analogy. 

In star formation studies SPH is popular because it is intrinsically Lagrangian. As opposed to mesh-based methods, it does not require a fixed grid to represent fluid properties and calculate spatial derivatives \citep{hockney88}. The fluid particles are free to move and -- in analogy -- constitute their own grid. The method is therefore able to resolve very high density contrasts, by increasing the particle concentration where needed. This it most effective, if the smoothing length is adaptable \citep{monaghan05}. There is no need for the complex and time-consuming issue of adaptive grid-refinement. However, the method has its weaknesses compared to grid-based methods. For example, its convergence behavior is mathematically difficult to assess and the method has problems reproducing certain types of dynamical instabilities \citep{Agertz07}. The algorithmic simplicity of the method and the high flexibility due to its Lagrangian nature, however, usually outweigh these drawbacks and SPH remains one of the numerical workhorses of current star-formation studies. There are various implementation of the method. Examples of very popular codes are GADGET \citep{springel01a,springel05a}, GASOLINE \citep{wadsley04}, MAGMA \citep{rosswog07a}, and the various decendents of the SPH program originally developed by Benz \citep{benz90,bate95,bate97}.

\subsubsection{Sink Particles as Subgrid-Scale Models for Protostars}
\label{subsubsec:sinks}
A fundamental problem for modeling protostellar collapse and star formation are the enormous density contrasts that need to be covered. Regions of high density require small cell sizes in grid-based methods \citep{truelove97}, or equivalently, small-particle masses in SPH calculations \citep{bate97,whitworth98a}. In order to guarantee stability, every numerical scheme must resolve the traversal of sound waves across the minimum resolution element, i.e.\ either across one cell or across the smoothing kernel of individual SPH particles. This is the so called Courant Friedrich Lewy criterion. It causes the time integration stepsize to get smaller and smaller as the density increases.  As a consequence, a computation virtually grinds to a halt during gravitational collapse.

When modeling the build-up of entire clusters of stars, or even following the accretion of the bulk of a core's mass onto a single star, this problem clearly needs to be overcome. One way out is to introduce sink particles. Once the very center of a collapsing cloud cores exceeds a certain density threshold (usually several thousand times the mean density, or using a threshold based on the Jeans mass) it is replaced by one single particle which inherits the combined masses, linear and angular momentum of the volume it replaces and which has the ability to accrete further gas from the infalling envelope. This permits to follow the dynamical evolution of the system over many global free-fall times, however, at the cost of not being able to resolve the evolution at densities above the threshold value.  In a sense, sink particles introduce ``inner boundaries'' to the computational domain. They have been successfully introduced to grid-based \citep{krumholz04} as well as particle-based methods \citep{bate95,Jappsen05}.

Each sink particle defines a control volume with a fixed radius. It lies  typically between a few and a few hundred astronomical units, AU, depending on the specific goals of the calculation. For comparison, the radius of Earth's orbit per definition is exactly $1\,$AU. In most cases the sink radius is chosen such that the Jeans scale below the threshold density is sufficiently resolved \cite{truelove97,bate97}. There are, however, implementations where sink particles have radii equivalent to one cell only.
Because the interior of the control volume is not accessible, the physical interpretation is often very difficult and subject to debate. Usually sink particles are thought to represent individual protostars or dense binary systems. This is supported by detailed one-dimensional implicit radiation hydrodynamic calculations which demonstrate that a protostar will build up in the very center of the control volume about $10^3\,\mathrm{yr}$ after sink creation \citep{wuchterl01}  which will swallow most of the infalling material.

Protostellar collapse is accompanied by a substantial loss of specific angular momentum, even in the absence of magnetic braking \citep{jappsen04,fisher04}. Still, most of the matter that falls in will assemble in a protostellar disk. It is then transported inward by torques from magnetorotational and possibly gravitational instabilities \citep{shu90,laughlin94,bodenheimer95,papaloizou95,lin96,gammie01,lodato05,kratter06,kratter08a}.  With typical disk sizes of order of several hundred AU in simulations of the formation of star clusters, the control volume fully encloses both star and disk. Even in higher resolution calculations that focus on single cores, the control volume contains the inner part of the accretion disk. If low angular momentum material is accreted, the disk is stable and most of the material ends up in the central star. In this case, the disk simply acts as a buffer and smooths eventual accretion spikes. It will not delay or prevent the mass growth of the central star by much.  However, if material that falls into the control volume carries large specific angular momentum, then the mass load onto the disk is likely to be faster than inward transport. The disk grows large and may become gravitationally unstable and fragment. This may lead to the formation of a binary or higher-order multiple \citep{Bodenheimer00,kratter08a}.  Indeed a initial binary fraction of almost 100\% is consistent with observations of star clusters \citep{Kroupa1995a}.  To some degree this can be taken into account by introducing an appropriate scaling factor. In a cluster environment the protostellar disk may be truncated by  tidal interactions and loose matter \citep{clarke00, Adams2006}. The importance of this effect depends strongly on the stellar density of the cluster and its dynamical evolution.  Further uncertainty stems from the possible formation of O or B stars in the stellar cluster. Their intense UV radiation will trigger evaporation and gas removal, again limiting the fraction of sink particle mass that turns into stars. Similar holds for stellar winds and outflows. These feedback processes are discussed  in Section \ref{sec:feedback} below.

\section{The Importance of Feedback}
\label{sec:feedback}

\subsection{Feedback Processes}

Most star formation simulations to date have neglected the effects of feedback on the star formation process. While this is computationally simpler, it is clearly not physically correct, and its omission leads to a number of obvious differences between simulations and observations. For example, simulations without feedback produce star formation that is too rapid and efficient \citep{krumholz07a} and significantly overproduce brown dwarfs compared to observations \citep{offner08b}. To make progress the next generation of simulations will have to remedy this omission.

\subsubsection{Radiation Feedback}

We begin our consideration of feedback processes by examining the effects of radiation from young stars. It is convenient to distinguish three distinct types of radiative feedback on star formation. The dominant sources of radiation in forming star clusters are the young massive stars, which begin their lives accreting rapidly, producing high accretion luminosities from the initial infall onto their surfaces. Later on in their formation these stars radiate prodigiously via Kelvin-Helmholtz contraction and then nuclear burning. The first effect of the radiation they produce is on their immediate environs. Their starlight is absorbed by dust grains suspended in the circumstellar gas, exerting a pressure that opposes gravity. Second, as the radiation diffuses out of the dusty gas clouds around a massive star it heats the gas. This affects the process of fragmentation, and thus plays a role in determining the stellar mass function for all stars born in strongly irradiated environments. Third, once massive stars contract onto the main sequence, they become significant sources of ultraviolet radiation, which can dissociate molecules, ionize atoms, and drive strong shocks throughout the star-forming cloud. These processes can both inhibit and promote star formation.

\textit{Radiation Pressure in Massive Star Formation.} The first effect is perhaps the best studied, and has been the subject of several recent reviews \citep{beuther07a, zinnecker07, mckee07, krumholz08b}, so we skip over it relatively quickly. As early as the 1970s researchers considering the formation of massive stars realized a fundamental problem. The largest stars in nearby galaxies have masses $\sim 100-150$ $\msun$ \citep{bonanos04, rauw05, figer05}, and for these stars radiation pressure on electrons within the star is the dominant support mechanism. In effect, these stars are at their internal Eddington limit. However, the Thompson cross-section is smaller than the cross-section of dusty gas to stellar radiation reprocessed into the infrared by an order of magnitude. Thus if a massive star is at its Eddington limit with respect the ionized, dust-free gas in its interior, it must exceed the Eddington limit by an order of magnitude with respect to the dusty gas found in molecular clouds. How then is it possible for dusty gas to accrete and form a massive star, since the outward radiation force on the accreting material should be significantly stronger than the pull of gravity \citep{larson71,kahn74,wolfire87,zinnecker07}?

Analytic treatments of the problem suggest that the solution lies in the non-sphericity of the accretion process: if the dusty gas around a protostar is sufficiently opaque, it can collimate the radiation, reducing the radiation force over some fraction of the solid angle to the point where gravity is stronger and accretion can occur \citep{nakano89, nakano95, jijina96, krumholz05a}. Simulations appear to bear out this solution, at least preliminarily. Hydrodynamic simulations in two dimensions using a flux-limited diffusion approach (see below) are able to form stars up to 40 $\msun$ before radiation pressure reverses infall \citep{yorke02}, while three-dimensional simulations show no signs of a limit on the upper masses of stars imposed by radiation pressure \citep{krumholz05d, krumholz09a}. In both the 2-D and 3-D cases, radiation is strongly beamed toward the poles of an accretion disk, allowing gas to accrete through parts of the equatorial plane shielded by the disk. In 3-D, this self-shielding is further enhanced by the organization of the gas into opaque, dense filaments, while radiation escapes through optically thin channels. This effect appears to allow the formation of stars with no clear upper mass limit.

\textit{Radiation Heating and the IMF.} The second radiative effect is heating of the gas, with the resulting modification of the initial mass function. Increasing the gas temperature suppresses fragmentation, and the observed overproduction of brown dwarfs in isothermal simulations \citep{bate03} is at least in part due to their omission of radiative feedback \citep{matzner05}. Similarly, radiative feedback is a strong candidate solution to another mystery about massive star formation: why would $\sim 100$ $\msun$ of gas, a mass  that represents tens to hundreds of Jeans masses at the typical densities and temperatures of a molecular cloud that has not yet begun to collapse, ever collapse coherently rather than fragmenting into many objects \citep{dobbs05, bonnell06d}?  A possible answer is that the accretion luminosity produced by the collapse of a dense core in a massive star-forming region is sufficient to suppress a high level of fragmentation, converting a collapse that might have produces $\sim 100$ small stars into one that produces only a few massive ones \citep{krumholz06b, krumholz08a}. However, the overall importance of this process and its details are subject to ongoing debate, and clearly more work is required on this important subject.

\begin{figure*}
\includegraphics[width=\textwidth]{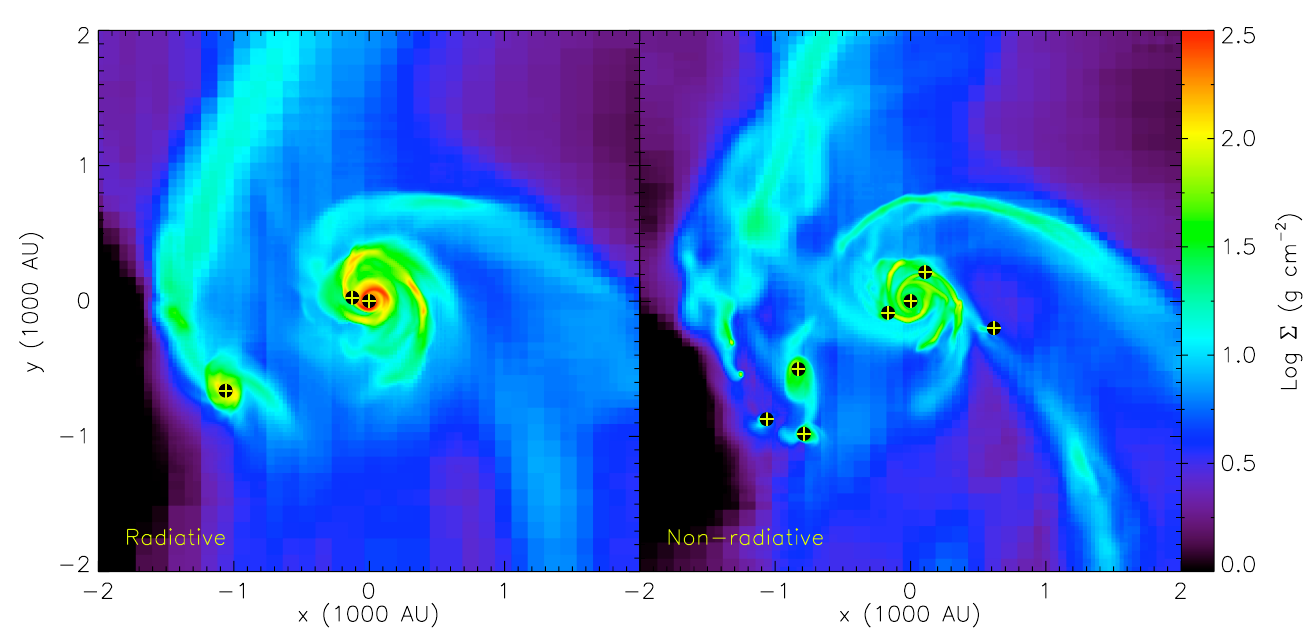}
\caption{\label{rtisocomp}
A comparison of two simulations with identical initial conditions and evolution times, one including radiative transfer (\textit{left panel}) and one done without it (\textit{right panel}). Stars are indicated by plus signs. The simulation without radiative transfer forms a factor of $\sim 4$ more stars than the one including it, and has significantly less mass in its gaseous disk. {\em (Images adapted from \citet{krumholz07a})}.
}
\end{figure*}

To date there is only one published simulation of the effect of radiative heating on fragmentation \citep{krumholz07a}, and it confirms the analytically-predicted outcome. Radiative heating reduces the amount of fragmentation that occurs during the collapse of massive pre-stellar cores. Figure \ref{rtisocomp} shows an example of this effect, comparing two simulations that are identical in every respect except that one is done with radiative transfer to one without it. However these simulations only studied single, isolated cores, and thus tell us little about the effect of radiative feedback on the IMF in star clusters. It remains unclear how radiative feedback in star clusters shapes the IMF. However it seems clear that, in the dense, optically-thick environments where clusters form, any results derived from simulations using the isothermal or optically-thin cooling approximations  will increasingly diverge from reality once high-mass stars begin to form.

\textit{High Energy Radiation and Star Formation Efficiency.} The third form of radiative feedback from massive stars is high energy radiation that is capable of dissociating hydrogen molecules (photon energies above $11$ eV)  and ionizing hydrogen atoms (photon energies above $13.6$ eV). The former creates a photodissociation region (PDR), a volume of mixed atomic and molecular gas at temperatures of hundreds of Kelvin, too warm to form stars. The latter rapidly heats the gas around a massive star to $\sim 10^4$ K and raises its sound speed to $\sim 10$ km s$^{-1}$, forming a structure known as an H~\textsc{ii} region. Except in the case of stars with very weak ionizing fluxes, or in environments where the magnetic field strongly confines the ionized region, the shock front generated by an expanding H~\textsc{ii} region generally overruns the PDR created by dissociating radiation and traps it between the ionization front and the shock front. For the purpose of molecular cloud dynamics, therefore, ionizing radiation is usually the more important effect \cite{krumholz07f}.

Unlike radiative heating and radiation pressure, dissociating and ionizing feedback do not become significant until fairly late in the star formation process. Early on rapid accretion swells massive stars to radii of $\sim 100$ $\rsun$ \citep{hosokawa08a, yorke08a}, and these large radii lead to low surface temperatures, reducing the fraction of a massive star's power that emerges at energies above $13.6$ eV. Moreover, even the full main sequence ionizing luminosity from a massive star will not escape from the stellar vicinity if accretion onto the star is sufficiently rapid and covers enough of the stellar surface \citep{walmsley95, keto02, keto03, tan03a}. 
In this quasi-spherical case, the H~\textsc{ii} region is kept from expanding and the Str\"omgren radius is small. 
However, the situation may change once protostellar outflows are taken into account (see next Section \ref{subsubsec:outflows}). These effectively remove high-density material along the rotational axis of the system. 
This may lead to an H~\textsc{ii} region that escapes along the outflow axis while remaining confined in the equatorial direction \cite{tan03a}, or it may allow the H~\textsc{ii} region the break free entirely from its parent core \citep{keto07}.

Once ionization does begin to break out of a massive protostellar core, it is likely to be the most significant of the three types of radiative feedback. Since 10 km s$^{-1}$ is much greater than the escape speed from a molecular cloud under Milky Way conditions, ionized gas escapes from star-forming clouds into the ISM, reducing the amount of mass available for star formation and unbinding molecular clouds \citep{mckee97, williams97}. Furthermore, since 10 km s$^{-1}$ is much larger than the 
sound speed in the non-ionized molecular gas, once they form H\textsc{ii} regions expand dynamically, driving shocks into the neutral material. Analytic models suggest that this can both promote star formation, by sweeping up gas into sheets that subsequently fragment by gravitational instability \citep{elmegreen77a, whitworth94a}, and inhibit it, by driving turbulent motions \citep{matzner02, krumholz06d}. Clearly more work is required to determine which effect dominates.

Simulations of these processes are still quite primitive, and most have focused on small molecular clouds that are already in a process of free-fall collapse when the simulation begins. Within this limited context simulations have produced a number of qualitative conclusions. First, single ionizing sources at the molecular cloud centers do not easily unbind those clouds, even if they deposit an amount of energy larger than the cloud binding energy \citep{dale05, maclow07}. This is because in a cloud with a pre-existing density structures, most of the energy is deposited in low-density gas that freely escapes from the cloud, while the higher density material is largely unaffected. Thus, in this context the effects of ionization on reducing the star formation efficiency are modest. Second, ionization can drive significant velocity dispersions in neutral gas, possibly generating turbulence \citep{mellema06}. Third, ionization does sweep up material and promote collapse, 
but numerical simulations indicate that this effect may again be modest \citep{maclow07, dale07a, dale07b}. Much of the swept up gas in these calculations was already on its way to star formation due to gravity alone, and the compression produced by an H~\textsc{ii} region shock only modifies this slightly.

This work is only a beginning, and many questions remain. First, none of the simulations to date have included multiple ionizing sources that are simultaneously active, so that interactions between expanding H~\textsc{ii} region shells can promote both star formation and turbulence. Since massive stars form in clusters, however, multiple sources should be the rule rather than the exception. Second, the simulations have for the most part focused on small, tightly-bound proto-cluster gas clouds being ionized by rather small ionizing luminosities corresponding to single stars, rather than larger, lower density, more loosely bound molecular clouds subjected to the ionizing flux of an entire star cluster. The effects of ionizing radiation may be greater in the latter case than in the former. Finally, only one simulation of ionizing radiation feedback to date has included magnetic fields \citep{krumholz07f}, and only then in a very idealized context. Since magnetic fields can tie together high- and low-density regions of a cloud, they may significantly increase the effects of ionization feedback.

\subsubsection{Protostellar Outflow Feedback}
\label{subsubsec:outflows}

Outflows from young stars provide another significant source of feedback on local scales in star-forming regions. During the process of accretion onto young stars about  $\sim 10\%$ of the gas that reaches the inner accretion disk is ejected into a collimated wind that is launched at a speed comparable to the Keplerian speed close to  the stellar surface. Theoretical predictions \citep{Pelletier92,shu00,Konigl00,Pudritz03,anderson05a,Pudritz07} and observational data \citep{bontemps96a,richer00} agree very well on this value. Another quantity that is well constrained by observations is the net momentum flux of the material entrained in the outflow. It is typically $\dot{p} \sim 0.3 \dot{M} v_{\rm K}$, where $\dot{M}$ is the accretion rate onto the star plus disk and $v_{\rm K}$ is the Keplerian velocity at the stellar surface \citep{bontemps96a,richer00,matzner02}. Outflow momentum flux correlates well with source luminosity across a very wide range in luminosity $L$, suggesting that all protostars show a common wind launching mechanism independent of mass \citep{wu04}. Since the wind momentum flux is much greater than $L/c$, this mechanism is almost certainly hydromagnetic rather than radiative in nature. The two primary theories for this are the x-wind \citep{shu00} and the disk wind \citep{Konigl00, Pudritz03,Pudritz07}. Common to both models is the idea that matter gets loaded onto magnetic field lines and then accelerated outwards by centrifugal forces. From the standpoint of feedback on scales large compared to the accretion disk around the source, the details of how the wind is launched matter little. All magneto-centrifugal winds approach the same distribution of momentum flux per unit angle at large distances from the launching region \citep{matzner99b}.  On larger scales, on which the outflow interacts with ambient material in the core, the opening angle varies depending on mass and age of the protostar. Outflows from low-mass stars appear quite well collimated, and remain so up to roughly B stars \citep{richer00,beuther05,Arce07}. The opening angles of O star outflows are wider, but it is unclear if this widening is an inherent property of the outflow or a result of the interaction between the outflow and the ambient gas.

One important difference between outflow and radiation feedback is that outflow feedback is more democratic. The most massive stars in a cluster dominate its radiative output, because (except at the very highest stellar masses) luminosity is a very strong function of mass, and ionizing luminosity an even stronger one \citep{kippenhahn94}. The dependence of luminosity on mass is strong enough to overcome the relative dearth of massive stars compared to low-mass ones. For outflows the reverse is true. The total mass accretion rate onto all the stars in a cluster is necessarily dominated by the low-mass stars, since they comprise the bulk of the stellar mass once star formation is complete. The Keplerian velocity at a star's surface varies as $\sqrt{M/R}$, where $M$ and $R$ are the star's mass and radius, and this ratio is only a very weak function of mass for main sequence stars. Thus, we might expect low-mass stars near the peak of the IMF to dominate outflow feedback. This simple analysis neglects the effect that more massive stars have shorter Kelvin-Helmholtz times and thus reach smaller radii more rapidly than low-mass stars, giving them larger Keplerian velocities at earlier times. Including this effect in a more careful analysis suggests that each logarithmic bin in mass contributes roughly equally to the total amount of momentum injected into a cluster \citep{tan02}. This has two important consequences: first, it means that outflow feedback can be important even in small clumps that do not form massive stars. Second, it means that simulations of outflow feedback cannot focus exclusively on the most massive stars, but must instead consider all stars as sources.

Outflows can influence their immediate surroundings as well as the cluster in which they form. On small scales, they reduce the star formation efficiency by removing mass from a collapsing core, both directly and via material that the outflow entrains as it escapes the core. Analytic estimates suggest that this process removes $25-75\%$ of the mass in a core \citep{matzner00}, but this is highly uncertain since no simulations of the collapse of individual cores with outflows that are capable of evaluating this estimate have been published. 

Outflows could also modify the star formation process by removing mass from protocluster gas clumps and possibly by driving turbulence within them \citep{norman80, mckee89, matzner00, matzner07}. However, it must be noted that this process cannot be the main driver of turbulence on global molecular cloud scales as outflows typically have short length scales. Instead this turbulence is probably driven on scales comparable to or larger than typical cloud sizes  \citep{maclow00b,Ossenkopf02,ossenkopf08a,ossenkopf08b}. One possible candidate for the origin of this large-scale turbulence is convergent flows in the galactic disk (see Section \ref{subsubsec:MCform}) either driven by gravitational instability in the disk \citep{li05b,li06,dobbs08b}, by collisions between molecular clouds \citep{Tasker09a}, or caused by supernova explosions 
\citep{MacLow05,2007ApJ...665L..35D}. Another candidate is giant H~\textsc{ii} regions created by clustered star formation within clouds, which have size scales comparable to entire GMCs \cite{matzner02, krumholz06d}. In addition, there seems to be no difference between the measured turbulence content of cloud clumps that are still in the so-called dark phase, i.e.\ before star formation has set in, and cloud clumps that are already actively building up stars in their interior \citep{ossenkopf01a}. This indicates that, at least at birth, star-forming regions must have turbulent motions that were imprinted as part of the formation process. Conversely, however, observations show that high column density star-forming clumps within GMCs lie above the linewidth-size relation observed for GMCs as a whole \cite{plume97,shirley03, heyer08}. This suggests that their turbulence cannot be supplied from large scales motions within the parent GMCs. Either these regions are powered by gravitational collapse, in a scaled-down version of the scenario described in Section \ref{subsubsec:MCform}, or they are driven by internal sources. Outflows are a natural candidate for this, and the deviation from a simple powerlaw linewidth-size relation predicted by analytic models appears to be consistent with what is observed \cite{matzner07}.

Simulations of protostellar outflows to date fall into two categories. Local simulations focus on the interaction of a single outflow with an ambient medium at high resolution, while larger-scale simulations follow an entire gas clump and star cluster including multiple outflows, but at significantly lower resolution. Local simulations attempt to understand the driving of turbulence by single outflows in detail. However, interpretation of these results is difficult, since there is no simple way to separate ``turbulence" from the coherent motion caused by a single outflow. Different authors analyzing simulations in different ways have come to opposite conclusions, with some arguing that outflows cannot drive supersonic turbulence \citep{Banerjee07b}, while others conclude that it can \citep{cunningham06b, cunningham08a}.

\begin{figure}
\includegraphics[width=0.7\columnwidth]{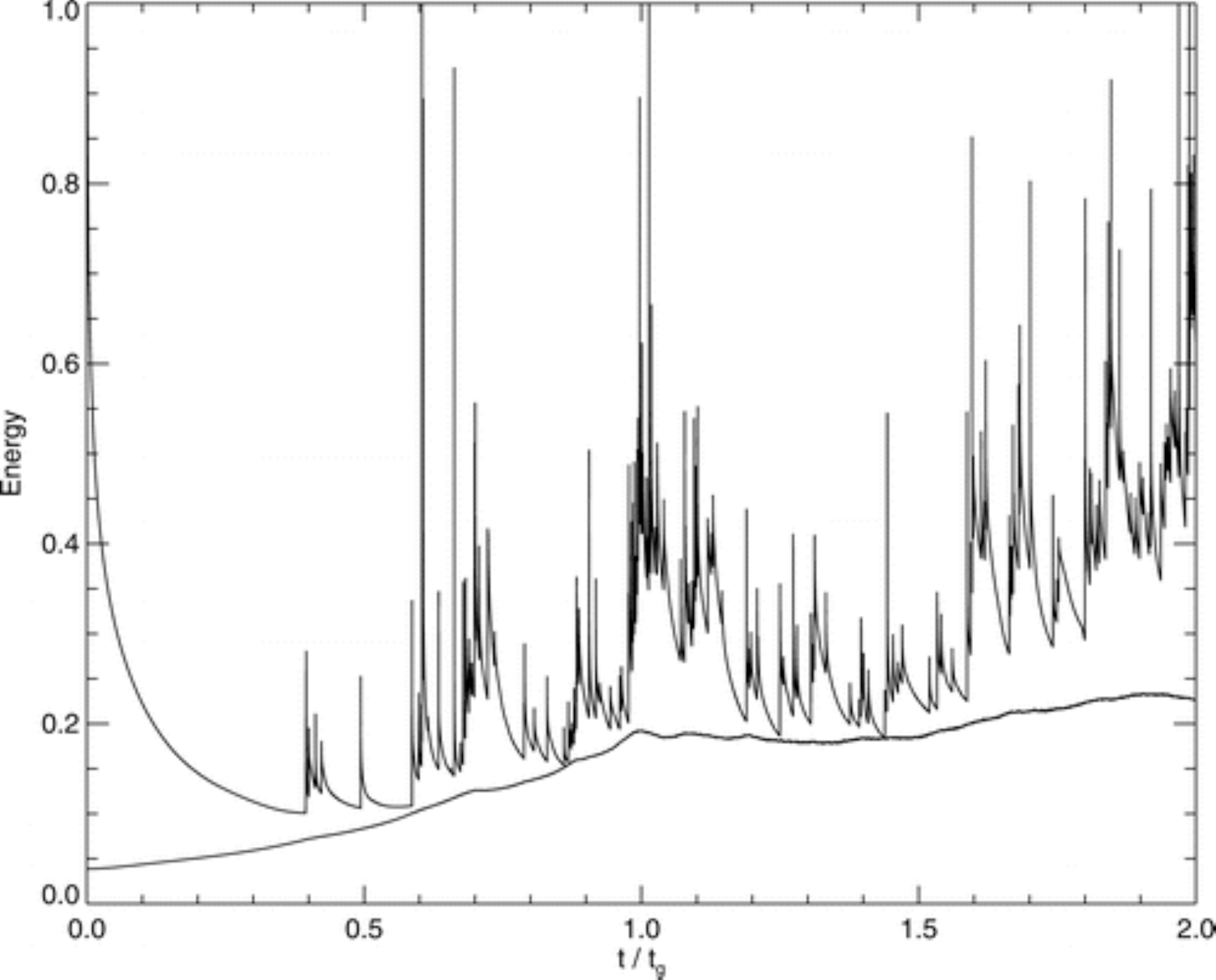}
\caption{\label{nakamurali}
Evolution of the total kinetic energy (\textit{upper line}) and gravitational energy (\textit{lower line}) as a function of time in a simulation of the formation of a star cluster including protostellar outflow feedback. Energies are normalized to the initial kinetic energy in the simulation, and times to the gravitational (or free-fall) time. {\em (Plot taken from \citet{nakamura07})}
}
\end{figure}

Global simulations of outflow feedback generally find that it has strong effects on the cluster formation process. In the absence of energy input, simulations of cluster-forming gas clumps find that any turbulence initially present decays rapidly, leading to a global collapse in which an appreciable fraction of the mass is converted into stars within a few dynamical times \citep{klessen98,klessen00a,klessen01,bate03,bonnell03}. Simulations that include outflow feedback found that outflows can change this picture. They eject mass from the densest and most actively star-forming parts of a cluster, reducing the star formation rate, while at the same time injecting enough energy to slow down overall collapse and maintain a constant level of turbulent motions \citep{li06b, nakamura07, nakamura08a}. As a result, the star formation rate drops to $<10\%$ of the mass being converted into stars per free-fall time, and rather than undergoing a runaway collapse the clump reaches a slowly evolving quasi-equilibrium state. Figure \ref{nakamurali} illustrates this effect in a simulation of the formation of a star cluster including protostellar outflow feedback. At the start of the calculation the kinetic energy falls as the initial turbulent velocity field decays. Consequently the cloud contracts and at about half of a free-fall time stars start to form and drive outflows. This energy input changes the subsequent evolution, and the cloud's global contraction is halted or at least significantly retarded. A definite answer would require to follow the dynamics over a longer period in time.

All the simulations published to date have significant limits. With only one exception they treat the wind as an instantaneous explosion, rather than a continuous beam injected over $\sim 10^5$ yr as we observe. The individual explosion spikes are clearly visible in Figure \ref{nakamurali}.  The current studies are also characterized by low numerical resolution ($128^3$ cells), which makes the energy injected more space filling, which is one of the main characteristics of interstellar turbulence. In addition, the calculations are performed in a periodic box, so energy cannot leave the star-forming cloud. In reality some very strong pencil-beam outflows escape their parent clumps and cover distances of a few parsec \citep{Stanke02,bally07a}. The one simulation published thus far that does include time history of accretion and better resolution only considers the effects of outflows from stars larger than 10 $\msun$ \citep{dale08a}, thereby neglecting the majority of the outflow power. Clearly the problem of outflow feedback and how it affects star formation is in need of further study.

\subsubsection{Other Types of Feedback}

Although radiation and protostellar outflows are thought to be the dominant feedback processes in star formation, two other mechanism are worthy of brief discussion: supernovae, and winds ejected by stars on the main sequence and post-main sequence. In terms of sheer energetics, it might seem odd to ignore supernovae as a major source of feedback. However, two compensating effects reduce their role in regulating star forming clouds. The first is timescales. Even the most massive stars do not explode as supernovae until $3-4$ Myr after formation \citep{parravano03}, and this is comparable to or longer than the formation time of star clusters. Thus supernovae come too late to affect the formation of individual star clusters, although they may be able to affect their parent giant molecular clouds, which have longer lifetimes.

A second effect, however, mitigates the impact of supernovae on GMC scales as well. Supernovae occur only after H\textsc{ii} regions and stellar winds have carved large cavities of hot, ionized gas around the massive stars that produce them. If a supernova occurs while this bubble is still embedded within its parent cloud, much of its energy is radiated away while the blast wave is still confined to the bubble. Simulations find that, as a result, the mass that is removed from the cloud by a supernova plus ionization is typically only $\sim 10\%$ larger than that removed by ionization alone \citep{tenoriotagle85,yorke89,matzner02,krumholz06d}. If, on the other hand, the bubble of hot gas created by ionization has broken out of a massive star's parentel cloud by the time the star explodes, both distance and an impedance mismatch make it difficult to deliver much of the supernova energy to the cloud. In a few Myr the expanding H\textsc{ii} region around a massive star cluster can push back the parent GMC by $\sim\,10$ pc or more from the site of the supernova, which then occurs in an ionized medium whose density is $2-4$ orders of magnitude lower than that of the cloud. Both the distance and the large density jump serve to shield a molecular cloud from the effects of a supernova, so that very little of the supernova energy is deposited in the molecular cloud. This effect means that, even on GMC scales, supernovae are unlikely to be the dominant feedback mechanism. It is important to note, however, that the energy that is not deposited in the molecular cloud itself does nevertheless affect the remainder of the ISM on galactic scales. Consequently supernovae are likely to dominate the energetics of the ISM on large scales \citep{maclow04}.

Main sequence winds are also thought to be subdominant as feedback mechanisms due to the effects of ionization \citep{mckee84, matzner02}. Stellar winds initially expand into a bubble of ionized gas created by the ionization from a massive star, and they create a radiative shock within that bubble where much of the wind energy is dissipated. Only after the stellar wind shock catches up to the ionization-created shock can stellar winds begin to provide feedback to the parent cloud. Even then the increase in total kinetic energy in the shock is modest for stars up to at least 35 $\msun$ \citep{freyer06a}, and even for $60$ $\msun$ stars is only of order unity \citep{freyer03a}.

\subsection{Modeling Feedback}
\label{subsec:modeling.feedback}

\subsubsection{Numerical Methods for Non-Ionizing Radiation Feedback}

Stars emit the bulk of their radiation in the visible part of the spectrum, but  the dusty clouds in which stars form are generally very opaque to visible light until late in the star formation process, when most of the gas has already been accreted or dispersed. As a result, direct stellar radiation tends to be absorbed by dust and reprocessed into the infrared close to the star that emits it, and modeling the resulting diffuse infrared radiation field is the primary goal of most numerical methods for simulating non-ionizing radiation feedback.

Focusing on the diffuse infrared radiation field simplifies the radiative transfer problem considerably, since the primary opacity source at infrared wavelengths is dust rather than atomic or molecular lines, and because in the IR scattering is negligible compared to absorption \citep{rybickilightman}. Even with these simplifications, though, it is possible to solve the full equations of (magneto-)hydrodynamics plus the equation of radiative transfer for this problem only in one dimension \cite{larson69, masunaga98, masunaga00}. Such an approach is unfortunately too computationally expensive to be feasible in three or even two dimensions. Instead, one must simplify the problem even further.

One approach is simply to modify the standard optically-thin cooling curve used in simulations without feedback by using an approximation to estimate the optical depth and reduce the cooling rate appropriately \cite{banerjee06b, stamatellos07b}. In this case one need not to solve a radiative transfer problem at all. This is an advantage, since it means that the radiation step is has nearly zero computational cost, but it is also a limitation. Because it lacks a treatment of radiative transfer, this approach allows gas to heat up due to adiabatic compression, but not because it is being illuminated by an external radiation source. In particular, in this approach there is no way for stars to heat gas. Since stellar radiation provides significantly more energy than gravitational compression once the first collapsed objects form \cite{yorke02,krumholz06b,krumholz07a,zinnecker07,krumholz08a}, this technique is only suitable for simulating star formation up to the point when the first parcels of gas collapse to stars.

The most common and simplest approach that can go past first collapse and follow either accretion or subsequent star formation, and the only one used so far in ``production" simulations \cite{yorke99, yorke02, krumholz05d, whitehouse06, krumholz07a}, is the flux-limited diffusion approximation \cite{alme73, levermore81}. The underlying  physical idea is simple: in an optically thick environment like the dusty clouds in which stars form, radiation diffuses through the gas like heat, and the radiative flux $\mathbf{F}$ obeys Fick's Law: $\mathbf{F} = -c \nabla E/(3\kappa \rho)$, where $E$ is the radiation energy density, $\rho$ is the gas density, and $\kappa$ is the specific opacity, with units of area divided by mass. This is the standard diffusion approximation, and it can be made either in a gray form by integrating $E$ and $\mathbf{F}$ over all frequencies, or in a multi-group form in which one divides the spectrum into some number of intervals in frequency and computes a separate energy density and flux for each interval \cite{yorke02, shestakov08}.

The pure diffusion approximation encounters a problem when the opacity is low, since if $\kappa\rho$ is sufficiently small the flux can exceed $cE$, violating the constraints of special relativity. The flux-limiting approach is to solve this problem by modifying the law for the flux to $\mathbf{F} = -\lambda c \nabla E/(\kappa \rho)$, where $\lambda$ is the flux limiter, a dimensionless function of $E$ and $\kappa \rho$ that has the properties $\lambda\rightarrow 1/3$ when the gas is optically thick, and $\lambda\rightarrow \kappa\rho E/|\nabla E|$ when it is optically thin. This limiting behavior ensures that flux approaches the correct Fick's Law value when the optical depth is high, and correctly reaches a maximum magnitude of $cE$ at low optical depth. Many functional forms are possible for $\lambda$. The most commonly-used one is the Levermore \& Pomraning limiter $\lambda = R^{-1} (\coth R - R^{-1})$, with $R=|\nabla E|/(\kappa \rho E)$ \cite{levermore81, levermore84}.

Given a formula for computing the radiation flux in terms of the radiation energy density, it is possible to drop all moments of the equation of radiative transfer except the zeroth one, so that the set of equations to be solved consists of the standard equations of HD or MHD, with some added terms describing the interaction of radiation with the gas, plus one additional equation for the radiation energy density. One treats feedback from stars in this formulation simply by adding it as a source term or a boundary condition in the radiation energy equation. The resulting set of equations may be written using either a comoving \cite{tscharnuter87a, boss92a, stone92c, whitehouse04, whitehouse05, hayes06} or a mixed-frame \cite{howell03, krumholz07b, shestakov08} formulation. The former approach is more suited to implementation in a code that is either Lagrangean, such as SPH, or based on van Leer advection \cite{van-leer77a}, but has the disadvantage that the equations are not explicitly conservative, and so the resulting codes cannot precisely conserve energy. The mixed-frame equations, on the other hand, are explicitly conservative, which makes them preferable for codes based on a conservative update, particularly those involving adaptive mesh refinement. In either form the equations are still significantly more expensive to solve than the corresponding non-radiative ones, since codes must handle radiative diffusion implicitly in order to avoid severe constraints on the time step, but solutions are within the reach of modern supercomputer simulations.

Although it is the tool of choice for star formation simulations at present, the pure flux-limited diffusion approximation does have some important limitations. Diffusion methods do not correctly represent shadowing effects which appear in systems that are optically thin or nearly so, nor can they model direct stellar radiation before it is absorbed and re-radiated isotropically. Pure diffusion also assumes that the gas and dust are thermally well-coupled.  While this approximation is a good one at densities $\sim 10^5$ cm$^{-3}$ or more, it may fail at lower densities. Diffusion also neglects cooling via molecular line emission, which can also be important at lower densities \citep{Genzel1991}.

The literature contains a variety of numerical techniques to address these shortfalls. One can handle imperfect dust-gas coupling by explicitly including it in the iterative radiative transfer update \cite{yorke99}. To handle direct stellar radiation or molecular cooling as well as the diffuse IR field, one can use a hybrid approach that combines a diffusion step with a ray-tracing step \cite{murray94a} or an optically-thin cooling step. To correctly model shadowing, one can use a more sophisticated radiative transfer method than diffusion, such as Monte Carlo, ray-tracing \cite{heinemann06a}, variable tensor Eddington factor (VTEF) \cite{hayes03}, or $S_n$ transport \cite{livne04a}. However, with the exception of the dust-gas coupling method, none of these techniques have thus far been used in any ``production" simulations of star formation. In some cases this is simply a matter of the necessary techniques not yet having been implemented into the codes most commonly used for star formation studies. These techniques, such as two-step approaches for the diffuse IR field and direct stellar fields and line radiation, are likely to appear in production simulations in the next few years. In other cases, however, the limitation is one of computational expense. For example the VTEF and $S_n$ methods have thus far only been used in two-dimensional calculations, simply because in three dimensions they have thus far proven prohibitively expensive. Remedying these problems will require significant advances in radiative transfer methodology to solve.

\subsubsection{Numerical Methods for Ionizing Radiation}

In comparison to non-ionizing radiation, handling ionizing radiation is conceptually more straightforward. Rather than a diffuse field arising from the repeated reprocessing of stellar radiation by dust grains, the ionizing radiation field in a star-forming region consists mostly of photons directly emitted from a stellar surface. Only in the outer parts of low-density ionized regions with sharp density gradients does reprocessed radiation make up a significant part of the photon field \cite{ritzerveld05}. The dominance of a relatively small number of point sources of radiation translates into a much simpler computational problem. By far the most common approach for solving it is to adopt the on-the-spot approximation \cite{osterbrock89}, in which one assumes that recombinations of ionized atoms into the ground state yield ionizing photons that are re-absorbed immediately near the point of emission. One therefore ignores photons emitted by recombining atoms entirely, and one solves the transfer equation along rays from the emitting stellar source, balancing recombinations into excited states against ionizations along each ray.

Within this over-arching framework, there are a variety of subtleties about how one draws the rays and updates the gas state. For example, the ray-drawing procedure can range in complexity from grids of rays restricted to radial paths originating at the center of a spherical grid \cite{whalen06}, up to a variety of schemes for handling casting rays either with a fixed \cite{garciasegura96, abel99a, mellema06, maclow07} or adaptive \cite{abel02b,rijkhorst05, krumholz07f} ray grid, or through a field of SPH particles \cite{kessel-deynet00a, dale07c}. Similarly, there are a variety of possible time-stepping strategies for handling the interaction of radiation heating with (magneto-)hydrodynamics. The simplest are Str\"omgren volume methods, in which one assumes that the gas reaches radiation and thermal equilibrium instantaneously \cite{garciasegura96, dale07c}. Solving time-dependent equations for the thermal and chemical structure but not resolving the relevant timescales hydrodynamically represents a middle ground \cite{kessel-deynet00a, mellema06, maclow07}, while the most complex option is to restrict the hydrodynamic time step to resolve gas heating and cooling times \cite{abel99a, whalen06, krumholz07a}. As always in numerics, there is a tradeoff between speed and quality of solution. Fully resolving the ionization heating time produces measurably more accurate solutions \cite{krumholz07f}, but is of course significantly more expensive than resolving it only marginally or assuming instantaneous equilibration.

\subsubsection{Numerical Methods for Protostellar Outflows}

Protostellar outflows are a natural result of the process of collapse and accretion that produces stars, and simulations of star formation that treat MHD and gravity with sufficient resolution do not need to include any additional physics to produce outflows \cite{banerjee06a, machida08c}. However, sufficient resolution here means that the simulation must resolve the outflow launching region, which is typically no more than $\sim 10$ stellar radii. Achieving such high resolution is prohibitively expensive for simulations that span more than a tiny fraction of the total formation time of a star, let alone an entire star cluster, so the most common procedure in such simulations is similar to that used for radiative feedback. Replace the collapsing region with a sink of some sort, and model an outflow emerging from that sink via a subgrid model that sits on top of whatever sink model the computation uses.

Such a model for outflows must specify the amount of mass and momentum contained, as well as the angular distribution of these quantities. Of these quantities the momentum is the best constrained by observations, since it remains unchanged even as the outflow gas entrains the material it encounters in the protostellar core. The mass flux and the angular distribution of the outflow are less well constrained, since these are altered as the outflow ages and interacts with its environment. The correct value to use in a given simulation probably depends on the length scale that simulation resolves, since the mass and opening angle both increase as ouflowing gas moves away from the star and interacts with its environment. Most simulations assume the standard value of 10\% for the ratio of infalling to outflowing mass. A good approximation for the opening angle is to assume the momentum of the outflowing gas is distributed with a profile $p\propto 1/(r\sin\theta)^2$, where $r$ is the distance from the star and $\theta$ is the angle relative to the star's axis of rotation \cite{matzner99b}. The opening angle adopted in numerical simulations, however, varies enormously all the way from $0^\circ$ \citep{Banerjee07b} to $90^\circ$  \citep{nakamura07}.

Once one has chosen a physical model for the outflow mass, momentum, and angular distribution in a given simulation, there remains the question of how to implement it numerically. As noted above, all stars contribute significant amounts of momentum, so any realistic approach must include contributions from any star that forms in a simulation. In grid codes this problem is generally straightforward; one simply adds mass and momentum with the desired angular distribution to the computational cells in or immediately around the sink region for each star \cite{nakamura07}. In particle-based codes the problem is more complex, since one must take care to avoid artificial clumping due to the discrete nature of the particles, and to ensure the momentum deposition in the region surrounding the sink is not altered by numerical interpenetration of the SPH particles. A variety of strategies are available to solve these problems \cite{dale08a}, but they are computationally expensive, which presents a potential problem for simulations with large numbers of sources.

Once the subgrid model is in place, outflows are much easier to simulate than radiation feedback, because outflow evolution is governed solely by hydrodynamics or MHD plus gravity, the physical mechanisms that are already included in any code used to simulate star formation. Beyond those involved in computing the subgrid model and injecting the outflow, the only additional computational cost that outflows impose on a code comes from the fact that outflow velocities can reach hundreds of km s$^{-1}$, significantly greater than the $\sil 10$ km s$^{-1}$ turbulent or infall speeds typically found in simulations that omit outflows. The higher speeds require smaller simulation time steps, and a corresponding increase in computational cost.

\section{Summary and Outlook}
\label{sec:outlook}

In this review we presented a overview of the current state of numerical star-formation studies. We have restricted ourselves to the early phases of stellar birth, from the formation of molecular clouds through to the build-up of stars and star clusters in their interior. We have left out the problem of accretion disks and protostellar evolution and point to other reviews in this context \citep{Hartmann1998,palla02b,StahlerPalla05}. 

We hope we have illustrated that the question of stellar birth in our Galaxy and elsewhere in the universe is far from being solved. Instead the field is rapidly evolving and has gone through a significant transformation in the last few years. In numerical star formation studies, we notice a general trend away from solely considering isolated processes and phenomena towards a more integrated multi-scale and multi-physics approach in todays computer simulations. In part this is triggered by the growing awareness that many physical processes contribute more or less equally to the formation of stars, such that it is not possible to single out individual effects. Reliable and quantitative predictions can only be made  on the basis of taking all relevant physical phenomena into account. Another reason for this development is the tremendous increase in computational capability provided by the advent of (relatively) easy-to-handle massively-parallel supercomputers, coupled with new and more efficient numerical algorithms for these machines.

If we examine the past and current state of the art, then it is evident that most studies so far have focussed on a small number of physical processes only. Typically, one had a single question in mind, such as what happens if we include one particular physical process? How does it affect the system? How does it modify possible equilibrium states? And how does it influence the dynamical evolution if we apply perturbations? The processes and phenomena about which these questions have been asked include hydrodynamics, turbulence, gravitational dynamics, magnetic fields, nonequilibrium chemistry, and the interaction of radiation with matter, but typically only one or two of them have been included in any given simulation. More sophisticated approaches include larger numbers of processes, but no simulation so far has considered all of them. The challenge in the past was mainly to do justice to the inherent multidimensionality of the considered problems. For example, stellar birth in turbulent interstellar gas clouds with highly complex spatial and kinematical structure is an intrinsically three-dimensional problem with one- or two-dimensional approaches  at best providing order-of-magnitude estimates. Including multiple physical processes gave way before the challenge of simulating in three dimensions.

This era is coming to an end. Many of today's most challenging problems are multi-physics, in the sense that they require the combination of many (if not all) of the above-mentioned processes, and multi-scale, in the sense that unresolvable microscopic processes can feed back onto macroscopic scales. This is true not only for star formation studies, but applies to virtually all fields of modern astrophysics. For example, the coagulation of dust species to larger particles or the interaction of dust with the radiation field from the central stars will eventually feedback into the dynamical behavior of the gas in protostellar accretion disks and hence has severe consequences for the formation and mass growths of planetary systems. Similarly, star formation and baryonic feedback are crucial ingredients of understanding galaxy formation and evolution in cosmological models. In a realistic description of cosmic phenomena, one is faced with the highly non-linear coupling between quite different kinds of interactions on a variety of scales. Star formation is no exception.

This is not only a challenge, it is also a chance, because it may open up new pathways to successful collaborations across astrophysical disciplines.   It also  reaches out to scientists in neighboring fields, such as applied mathematics or computer science. For example, only few groups around the world are able to fully benefit from the massively parallel computing architectures that are currently being developed. Peak performances with $\sim$ 100 teraflops will only be attainable on thousands of CPUs, sustained petaflop computing may require as many as $10^5$ CPUs. This asks for a completely new approach to parallel algorithm design, a field where modern computer science is far ahead of the schemes currently used in astrophysics and star-formation studies. Regular methodological exchange with applied mathematicians and possibly numerical fluid dynamicists thus holds the promise of both transferring new methods into astrophysics and raising the awareness of mathematicians about numerical challenges in astrophysics.

Towards  the end, we want to speculate about a few of the what we think are the most interesting and pressing open problems in modern star formation theory and where the current advancements in computational power and algorithmic sophistication are likely to have a major impact.

\textit{What drives interstellar turbulence?} Observations show that turbulence in molecular clouds is ubiquitous, and that, with the exception of the dense cores discussed above, it seems to follow a universal relationship between velocity dispersion and size. Even extragalactic molecular clouds appear to obey similar scalings. There are no variations in the turbulent properties between GMCs with little and much star formation, which might seem to argue for galaxy-scale driving, but there is also no systematic variation in GMC properties within a galaxy or between galaxies, which would seem to argue that internal processes must be important. So what is the relative importance of internal and external forcing mechanisms in driving ISM turbulence? Does the answer depend on the length scales that one examines, or on the place where one looks?

\textit{How does the multi-phase nature of the ISM influence stellar birth?} Star formation appears to follow fairly universal scaling laws in galaxies that range from mildly H\textsc{i}-dominated (such as the Milky Way) to galaxies that are strongly H$_2$-dominated (such as local starbursts). Does the presence or absence of a significant atomic phase play an important role in regulating star formation, either directly (e.g.\ by limiting the amount of molecular gas ``eligible" for star formation) or indirectly (e.g.\ by driving turbulent motions via thermal instability)? How does the star formation process change, if at all, in galaxies such as dwarfs that contain very little molecular gas?

\textit{How does stellar feedback influence star formation?} Stars produce a wide variety of feedbacks: outflows, main sequence winds, ionizing and non-ionizing radiation, and supernovae. Which, if any of these, are responsible for controlling the rate and efficiency of star formation? Does the answer to this question change in different galactic environments, i.e.\ are there different processes acting in the denser molecular clouds found in circum-nuclear starbursts than in the tenuous outer regions of the galaxy?

\textit{What determines the statistical properties of a stellar population, and are these properties universal?} On the observational side, is the stellar IMF and binary distribution at present days different in different galactic environments, or is it truly universal? Especially in rich clusters our observational basis still needs to be extended. The same holds for variations with metallicity as can be traced in the Local Group. Is the IMF in the Large Magellanic Cloud (with metal abundances of $\sim\,$$1/3$ of the solar value) and the Small Magellanic cloud (with $\sim\,$$1/10$ of that value) really similar to the Milky Way? On the theoretical side, what processes are responsible for the (non-)variation of the IMF? The critical mass for gravitational collapse can vary enormously between different environments. Yet the IMF in  globular clusters, for example,  appears to be the same as in regions of distributed star formation as in Taurus.  Hence, there must be additional physical processes that influence the fragmentation behavior of the interstellar gas and determine the resulting stellar mass spectrum.

\acknowledgements{We thank all our collaborators for many stimulating, encouraging and sometimes controversial discussions. In particular, we thank Javier Ballesteros-Paredes, Robi Banerjee, Ian A. Bonnell, Paul C.\ Clark, Bruce G.\ Elmegreen, Simon C.\ O.\ Glover, Lee W.\ Hartmann, Patrick Hennebelle, Anna-Katharina Jappsen, Richard I.\ Klein, Kaitlin M.\ Kratter, Mordecai-Mark Mac~Low, Christopher D.\ Matzner, Christopher F.\ McKee, Stella S.~R. Offner, Jonathan C.\ Tan, Todd A.\ Thompson, and Enrique V{\'a}zquez-Semadeni.
RSK thanks for support from the Germany Science Foundation, {\em DFG}, via the Emmy Noether grant KL 1358/1, grants KL 1358/4 and KL 1358/5, and the priority program SFB 439 {\em Galaxies in the Early Universe}.
MRK acknowledges support from National Science Foundation grant AST-0807739 and from the National Aeronautics and Space Administration through the Spitzer Space Telescope Theoretical Research Program, through a contract issued by the Jet Propulsion Laboratory, California Institute of Technology.
FH acknowledges support from National Science Foundation grant AST-0807305 and from the Natinoal Aeronautics and Space Administration through the Herschel Cycle-0 Theoretical and Laboratory Astrophysics Research program, NHSC 1008.
}

%\bibliographystyle{apsrev}
%\bibliographystyle{apsrmp}
%\bibliography{aslrefs}

\end{document}